\documentclass[preprint,groupedaddress,floatfix,showpacs,showkeys]{revtex4-1}
\usepackage[utf8]{inputenc}
\usepackage{amsmath}
\usepackage{amssymb}
\usepackage[mathlines]{lineno}
\usepackage{graphicx}
\usepackage[american]{babel}
\usepackage[amssymb]{SIunits}
\usepackage{subfigure}
\usepackage{notoccite}
\usepackage{afterpage}

\usepackage{caption}
\usepackage{adjustbox}
\usepackage{array}
\usepackage{bm}

\begin{document}

\title{Phase-field modeling of equilibrium precipitate shapes under the influence of coherency stresses}

\author{Bhalchandra Bhadak}
\email{Corresponding author:bhalchandrab@iisc.ac.in}
\affiliation{Department of Materials Engineering, Indian Institute of Science, 560012 Bangalore, India}

\author{R. Sankarasubramanian}
\email{sankara@dmrl.drdo.in}
\affiliation{Defence Metallurgical Research Laboratory, Hyderabad - 500058, India}

\author{Abhik Choudhury}
\email{abhiknc@.iisc.ac.in}
\affiliation{Department of Materials Engineering, Indian Institute of Science, 560012 Bangalore, India}


\begin{abstract}
Coherency misfit stresses and their related anisotropies are known to influence the equilibrium 
shapes of precipitates. Additionally, mechanical properties of the alloys are also dependent on the 
shapes of the precipitates. Therefore, in order to investigate the mechanical response of 
a material which undergoes precipitation during heat treatment, it is important to derive 
the range of precipitate shapes that evolve. In this regard, several studies have been 
conducted in the past using sharp interface approaches where the influence of elastic energy 
anisotropy on the precipitate shapes has been investigated. 
In this paper we propose a diffuse interface approach which allows us to minimize grid-anisotropy 
related issues applicable in sharp-interface methods. In this context, we introduce a novel 
phase-field method where we minimize the functional consisting of the elastic and surface 
energy contributions while preserving the precipitate volume. Using this method 
we reproduce the shape-bifurcation diagrams for the cases of pure dilatational misfit 
that have been studied previously using sharp interface methods and then extend them 
to include interfacial energy anisotropy with different anisotropy strengths which has 
not been a part of previous sharp-interface models. 
While we restrict ourselves to cubic anisotropies in both elastic and interfacial energies in this study, 
the model is generic enough to handle any combination of anisotropies in both the bulk and interfacial terms. 
Further, we have examined the influence of asymmetry in dilatational misfit strains along 
with interfacial energy anisotropy on precipitate morphologies.
\end{abstract}

\maketitle

\section{Introduction}

Precipitate strengthened alloys are one of the most commonly used materials 
for high temperature applications, whereby, the strengthening is achieved 
through precipitate-dislocation interactions. The mechanical properties 
of these alloys predominantly depend on precipitate size, morphologies and their 
distribution. Thus, there are several experimental studies carried out which 
focus on the precipitate morphology~\cite{conley1989}-~\cite{maheshwari1992}, 
their growth and coarsening ~\cite{ardell1970}-~\cite{rastogi1971}-
~\cite{chellman1974}-~\cite{ardell1985}, 
strengthening~\cite{seidman2002}-~\cite{pollock2006}. 
In this regard, there are two mechanisms, one in which the precipitates 
are large enough such that there is no coherency between 
the precipitate and the matrix from which it formed, 
and the second in which the precipitates are small such that 
there is still substantial coherency between the matrix and the precipitate. 
While in the former, the interaction of the dislocation with the precipitate 
is purely physical, in the latter, the coherency stresses around the precipitate 
also influence the interaction with the precipitate. In both cases the shape 
of the precipitate plays an important role in the interaction with the dislocation. 
Our investigation in this paper, will be related to the investigation of shapes 
of coherent precipitates, more particularly the understanding of the equilibrium 
morphology of precipitates as a function of the misfits, elastic and interfacial 
energy anisotropies. 

The first theoretical efforts are from 
Johnson and Cahn ~\cite{johnson1984elastically} who predict an equilibrium shape 
transition of elastically isotropic misfitting precipitate in a stiffer matrix.  
The equilibrium shape of a precipitate is determined by minimization under
the constraint of constant volume of the precipitate of the total energy, 
constituting of the sum of elastic and interfacial energies. The theory 
proposes the shape transition with size, akin to a second-order phase transition 
with the shape of the precipitate as an order parameter. The theory analytically 
predicts the equilibrium shape order parameters as a function of precipitate 
size whereby below a critical size the equilibrium shape is purely cylindrical 
with circular cross-section, 
whereas, beyond it the shape order parameter smoothly bifurcates into two 
variants. In their work, one of the transitions the authors discuss is 
the case where, beyond a particular size 
of a precipitate, a purely circular cross-section of the cylindrical precipitate 
elongates along either of two-orthogonal directions, thereby retaining only 
a two-fold symmetry. This is therefore, termed as a symmetry breaking transition. 

Voorhees and coworkers~\cite{voorhees1992,voorhees_su_thompson1994} 
give numerical predictions of equilibrium morphologies of 
precipitate with dilatational and tetragonal misfit in elastically anisotropic medium 
(cubic anisotropy). In this work, the authors discretize the interface co-ordinates
in terms of the arc-length, and use this to write the total energy of the system, 
which is a sum of the elastic and the interfacial energy.  Thereafter, the minimizer of
the total integrated energy of the system is described by the one which satisfies
the jump condition at the interface. The condition equates the departure of the chemical potential 
at the interface to the sum total of the jumps in the elastic configuration force and 
the capillarity forces. This balance condition for each of the discretized interface points
leads to a set of integro-differential equations, which along with the constraint of 
constant area, leads to the determination of the constant chemical potential 
that satisfies the interface balance conditions at 
all the co-ordinates. The solution set for the co-ordinates thereby constitutes
the equilibrium shape of the particle. The method has also been extended into 
three dimensions in the work by Voorhees and co-workers ~\cite{thompson1999, voorhees_li_lowengrub2004}.

Schmidt and Gross ~\cite{schmidt1997} use a boundary integral technique 
to perform the energy minimization in order to determine the equilibrium precipitate shape 
in elastically inhomogeneous and anisotropic medium, which is also used for studying 
multi-particle interactions in \cite{schmidt_mueller_gross1998}.
As an extension Schmidt et al.~\cite{mueller1998}, have determined the 3D equilibrium precipitate morphologies
again using a boundary integral method. 
They have observed that, there is a strong influence of elastic inhomogeneity, elastic anisotropy 
and characteristic length on the morphological stability of precipitates. 
Similarly, Jog et al. ~\cite{jog2000symmetry},~\cite{Sankarasubramanian2002} 
use a finite element method coupled with a congruent-gradient based optimization technique to minimize the sum of 
elastic and interfacial energies to determine equilibrium precipitate shape of isolated, 
coherent particle with cubic anisotropy in elastic energy. They have also estimated 
symmetry breaking as a function of particle size
for different combinations of elastic anisotropies and misfits.
Jou et al.\cite{lowengrub_leo_jou1997} use a boundary integral method to study 
single precipitate growth as well two-precipitate interactions.
Kolling et al.~\cite{kolling2003}, use a 
finite element and optimization based method to predict equilibrium precipitate shape 
for misfitting particle with dilatational eigenstrain. 
The authors show that the value of the elastic constant, 
particle size and inhomogeneity affect the stability of equilibrium shape 
of precipitate. Akaiwa et al. \cite{voorhees_akaiwa2001}
utilize the boundary intergal method with a fast multipole method for
integration to simulate growth and coarsening of large 
number of precipitates. 

An interesting extension of the finite element based method for the 
prediction of the equilibrium shape is the utilization of the level-set
method where a prescribed value of the level-set surface is used to 
describe the interface morphology \cite{voorhees_duddu2011}. The model 
is used for modeling solid-state dendrite growth as well coarsening by 
coupling to the composition field.
The modification of the modeling scheme to study equilibrium shape 
of precipitates is done by using a Lagrange parameter which 
maintains the same volume of the precipitate while the normal 
velocities of the interfacial nodes is calculated \cite{zhao_duddu_bardas2013}, 
where the composition field is treated uniform. 
The equilibrium shape is determined when the velocity of all the interfacial nodes
becomes zero. As an extension of the model, the authors also 
study chemo-mechanical equilibrium where the misfit strains are 
a function of the composition in \cite{zhao_bardas2015}, where diffusional equilibrium 
is solved along with the mechanical equilibrium.

In terms of the determination of the equilibrium morphologies of precipitates
the preceding models are all sharp interface-based methods and in addition
prescribe optimization solutions for the shape of the equilibrium precipitate 
under the constraint of constant volume. A corresponding approach using an 
microstructural evolution based method is proposed by Khachaturyan et al.~\cite{wang1991}, 
wherein, precipitate evolution is simulated using a modified
version of Cahn-Hilliard equation~\cite{cahn1958} until
there is complete loss of saturation or the situation where the 
diffusion potential everywhere becomes the same. 
In this situation, the shape of the precipitate, corresponds to equilibrium. However, 
there is no explicit volume constraint during evolution and the variation 
of the equilibrated shapes as a function of the sizes/volume of the precipitate
may only be tracked through a corresponding change in the 
alloy composition, yielding different volume fractions of the 
precipitate in a box of given size. Using this model the authors 
have studied effect of elastic stresses on precipitate shape instability during 
growth of single precipitate embedded in elastically anisotropic system.
Similar models have also been formulated by Leo et al. ~\cite{lowengrub_leo1998},
where the authors solve for diffusional equilibrium along with mechanical 
equilibrium in order to determine the equilibrium particle shape.
The authors also compare with the results from their corresponding
boundary-integral based modeling methods \cite{lowengrub_leo_jou1997}. 
The authors also highlight
the advantage of using the diffuse-interface schemes where merging
and splitting of particles can be handled easily in contrast to the 
sharp-interface methods which require re-meshing.
A similar dynamic evolution model has also been used to predict 
equilibrium shapes in three dimensions in ~\cite{voorhees_jokisaari2017}.
Apart from the determination of equilibrium shapes the diffuse-interface 
methods have also been used for the study of growth and coarsening
of multi-particle systems 
\cite{lowengrub_leo1998, wang_zhu_ardell_chen2004,chen_twang2008,tsukada2009,kundin2012,kundin_wang_mushongera2015, finel_cottura2015}, 
instabilities \cite{wang_Khachaturyan1995} 
and rafting \cite{gururajan2007,tsukada2008,gaubert_finel2010,boussinot_finel2010,wang_zhou2010, tsukada2011}.

In all the above studies, the central focus of investigation has been the 
study of equilibrium shapes of precipitates where the anisotropy exists only in the 
elastic energy. The coupled influence of both the interfacial energy and elastic 
anisotropies on the equilibrium morphologies 
is performed using the boundary integral method by Leo et al.
\cite{lowengrub_leo2000}.
In a study by Greenwood et al.~\cite{greenwood2009}, 
the authors have developed a phase field model of microstructural evolution, where 
they study the morphological evolution of solid-state dendrites as function of 
anisotropies in both surface as well as elastic energy. While, the study is not particularly 
aimed at the study of computing equilibrium shapes, the authors  illustrate transitions 
in the growth directions of solid-state dendrites from those dominated by the surface 
energy anisotropy to those along elastically soft directions, stimulated by a change in the 
relative strengths of the anisotropies. 

This brings us to the motivation of our paper, wherein, we formulate
a diffuse-interface model for the determination of equilibrium shapes of
precipitates under the combined influence of anisotropies in both elastic 
as well as interfacial energies, with any combination of misfits. 
Here, the shape of the precipitate will be defined by a smooth function, 
which varies from a fixed value in the bulk of the precipitate to another 
in the matrix. The shape of the precipitate will be described by a fixed 
contour line of this function. Our work can be seen as a 
diffuse-interface counterpart of previous work by Voorhees et al.
~\cite{voorhees1992} and other sharp-interface models by 
Schmidt and Gross ~\cite{schmidt1997} and Jog et al.~\cite{jog2000symmetry}
as well as the level-set based FEM method proposed in Duddu et al.\cite {voorhees_duddu2011}
and Zhao et al.\cite{zhao_duddu_bardas2013,zhao_bardas2015}.
Herein, we use a modification of the volume-preserved Allen-Cahn evolution 
equation that was proposed previously by Nestler et al.\cite{garcke2008}, 
wherein the Allen-Cahn equation prescribing the evolution of a given 
parameter is modified such that the integrated change in the volume 
that is computed over the entire domain of integration returns zero. The motivation for 
proposing a diffuse-interface model is three-fold: firstly given the 
Allen-Cahn dynamics of order parameter, which ensures energy minimization, 
there is no requirement for an additional optimization routine that is 
used in corresponding works by Schmidt and Gross ~\cite{schmidt1997} 
and Jog et al.~\cite{jog2000symmetry}. Secondly, complicated discretization 
and solution routines adopted by ~\cite{voorhees1992} can be 
avoided allowing for easy extension from two dimensions (2D) to three dimensions (3D). 
Thirdly, a diffuse-interface approach allows for an easy coupling of elastic and 
surface energy anisotropies.
Further, a corollary to the ease of discretization is that quite complicated
shapes with rapid change in curvatures can be treated with a great deal of 
accuracy , which we will demonstrate by comparing our results with those 
from a sharp interface model. 
Additionally, our diffuse-interface presentation will be different
from the previous phase-field models where we do not solve for the composition field 
while deriving the equilibrium shape of the precipitate 
which allows us to reduce the computational effort. 

In the following sections, we firstly describe the model, followed by a discussion 
on the results which include benchmarks with analytical as well as with a previous 
FEM based model by Jog et al.~\cite{jog2000symmetry}. We then continue with 
other combinations of tetragonal misfits and elastic anisotropies, finally 
concluding with a study involving a competition between elastic and interfacial 
energy anisotropy and its influence on equilibrium shapes. We conclude, with 
comparative statements between the different approaches and possible extensions 
of the model for investigations in multi-component systems.

\section{Model formulation}

During solid-state phase transformations, a difference of the lattice parameter between 
the precipitate and the matrix gives rise to misfit strains/stresses 
for a coherent interface. This in turn, contributes to the system energy in terms of
an elastic contribution which scales with the volume of the precipitate. Similarly, 
the interfacial energy which is the other component of the energy in the system, 
varies with the interfacial area. In this context, the equilibrium shape of 
the precipitate is the one which minimizes the contributions 
of both the elastic energy and interfacial components, 
which given the scaling of the two energy components is 
a function of the size of the precipitate.

In this section, we formulate a phase-field model, where the functional 
consists of both the elastic and the interfacial energy contributions.
Since, the equilibrium precipitate shape depends on the size of the 
precipitate, we formulate a model which minimizes the system 
energy while preserving the volume of 
the precipitate, and thereby allows the computation of the equilibrium 
shape of the precipitate shapes. This allows, the determination of the precipitate
shapes as a function of the different sizes as has been done 
previously using sharp-interface methods. 
This constrained minimization is achieved through 
the technique of volume preservation which is also described 
elsewhere ~\cite{garcke2008}, that is 
essentially the coupling of the Allen-Cahn type evolution
with a correction term using a Lagrange parameter that 
ensures the conservation of the precipitate volume during 
evolution.

In the following we discuss the details of the phase-field model.
We write the free energy functional as, 

\begin{align}
   \cal F(\phi) &= \int_V \Bigg[\gamma W a^{2}\left(\bm{n}\right)|\nabla \phi|^{2} 
		   + \dfrac{1}{W} w\left(\phi\right)\Bigg] dV \\ \nonumber
		  &+ \int_V f_{el}\left(\bm{u},\phi\right)dV + \lambda_\beta \int_V h\left(\phi\right)dV,
\end{align}

where V is the total volume of system. 
$\phi$ is the phase field order parameter that describes the presence and absence of
precipitate ($\alpha$ phase) $\phi=1$ and matrix ($\beta$ phase) $\phi=0$ phases.
The first integral constitutes the interfacial energy contribution, 
which in a phase-field formulation is done using a combination of the gradient-energy 
and potential contributions. Here, the term $\gamma$ controls the interfacial energy 
in the system, while $W$ influences the diffuse interface width separating the 
precipitate and the matrix phases. The function  $a\left(\bm{n}\right)$ which is a
function of the interface normal $\bm{n}=-\dfrac{\nabla \phi}{|\nabla \phi|}$, describes
the interfacial energy anisotropy of the precipitate/matrix interface. 
The second term in first integral describes the double well potential, 
which writes as,
\begin{align}
    w\left(\phi\right) &= \dfrac{16}{\pi^2} \gamma \phi\left(1-\phi\right) \qquad  \phi \in [0,1],\\ \nonumber
                           &= \infty \qquad \qquad \qquad \textrm{otherwise}.
\end{align}
The second integral enumerates the elastic energy contribution 
to the free-energy density of the system which is a function of the order parameter $\phi$
that is also used to interpolate between the phase properties and the misfit
as well as the displacement field $\bm{u}$.
$\lambda_\beta$ is the Lagrange parameter that is added in order that the 
volume of the precipitate represented by the integral $\int_V h\left(\phi\right)dV$, 
where $h\left(\phi\right)$ is an interpolation polynomial that smoothly varies 
from 0 to 1, is conserved. The evolution of the order parameter $\phi$ is derived using the 
classical Allen-Cahn dynamics that writes as, 

\begin{align}
 \tau W \dfrac{\partial \phi}{\partial t} &= - \dfrac{\delta F}{\delta \phi},
\end{align}

that elaborates as, 

 \begin{align}
  \tau W \frac{\partial\phi}{\partial t} &= 2 \gamma W \nabla \cdot \left[a\left(\bm{n}\right)\left[\dfrac{\partial a\left(\bm{n}\right)}{\partial \nabla \phi}|\nabla \phi|^{2} + a\left(\bm{n}\right) \nabla \phi\right]\right] \nonumber\\
                                                &- \dfrac{16}{\pi^2} \dfrac{\gamma}{W} \left(1-2\phi\right) - \dfrac{\partial f_{el}\left(\bm{u},\phi\right)}{\partial \phi} - \lambda_{\beta} h^{\prime}\left(\phi\right),
                                                \label{phi_evolve}
 \end{align}
 
 where $\tau$ is the interface relaxation constant, which in the present modeling 
 context is chosen as the largest value that allows for a stable explicit temporal evolution 
 using a simple finite difference implementation of the forward Euler-scheme.
 In order to complete the energetic description, it is important to elaborate the 
 elastic energy density $f_{el}\left(\bm{u}, \phi\right)$ in terms of the physical 
 properties of the matrix and the precipitate phases, that are the stiffness matrices, 
 as well as the misfit. Here, we adopt two ways of interpolating the phase properties, 
 one of them bearing similarity to the Khachaturyan scheme ~\cite{khachaturyan1983} of 
 interpolation and the other which is a tensorial interpolation which is elaborately 
 described in ~\cite{schneider2015}.
 
 \subsection{Khachaturyan interpolation}
In this interpolation method, the elastic energy density writes as,
\begin{align}
f_{el}(\bm{u}, \phi) &= \dfrac{1}{2}C_{ijkl}(\phi)(\epsilon_{ij} - \epsilon^*_{ij}(\phi))(\epsilon_{kl} 
- \epsilon^*_{kl}(\phi)),
\end{align} 
where the total strain can be computed from the displacement $\bm{u}$ as,
\begin{align}
 \epsilon_{ij} &= \frac{1}{2}\left(\frac{\partial u_i}{\partial x_j} + \frac{\partial u_j}{\partial x_i}\right)
 \label{strain},
\end{align}
while the elastic constants $C_{ijkl}$ and eigenstrain $\epsilon^*_{ij}$ can be expressed as:
\begin{align}
 C_{ijkl}(\phi) &= C^{\alpha}_{ijkl}\phi + C^{\beta}_{ijkl}(1-\phi), \\ \nonumber
 \epsilon^*_{ij}(\phi) &= \epsilon^{* \alpha}_{ij}\phi + \epsilon^{* \beta}_{ij}(1-\phi). 
\end{align}
To simplify the equations, without any loss of generality we additionally impose that
the eigenstrain exists only in precipitate phase ($\alpha$ phase), which makes $\epsilon^{* \beta}_{ij}=0$.
Thereafter, the elastic energy density can be recast as,  
\begin{align}
 f_{el}(\phi) &= Z_3 (\phi)^3 + Z_2 (\phi)^2 + Z_1 \phi + Z_0,
 \label{fel_phi}
\end{align}
where, in Eqn.~\ref{fel_phi}, we segregate the terms in powers of $\phi$, 
i.e. Z3, Z2, Z1, Z0. Each pre-factor is a polynomial of $\phi$, elastic constants and the 
misfit strains. The expansion of these pre-factors is illustrated in appendix.
Therefore, the term corresponding to the elastic energy in the evolution 
equation of the order parameter can be computed as, 

\begin{align}
 \dfrac{\partial f_{el}\left(\bm{u},\phi\right)}{\partial \phi} &=  3Z_3 (\phi)^2 + 2Z_2 (\phi) + Z_1. 
\end{align}

\subsection{Tensorial interpolation}
\label{Tensorial}
In this subsection, we utilize an interpolation scheme which allows the stress/strain terms
normal and tangential to the phase-interface to be interpolated differently. An elaborate description 
for the motivation behind this can be found in ~\cite{schneider2015,durga2013}. 
Concisely, this allows for an efficient control on the excess energy at the interface that in 
principle should enable an easier comparison with analytical models. We elaborate the scheme 
for two-dimensions, but in principle can also be generalized for more than two dimensions.

In this interpolation scheme, we rotate the stiffness tensors which are prescribed in the 
Cartesian co-ordinate system into the system that is described by the local unit normal vector
and the tangent of the $\alpha/\beta$ interface. This is done using the unit normal vector of the 
interface, 
\begin{equation}
 \bm{n} = -\dfrac{\nabla \phi}{|\nabla \phi|}.
\end{equation}

Describing a transformation from the Cartesian system $\bm{x}$, $\bm{y}$ to $\bm{n}$, $\bm{t}$ where after rotation of $\bm{x}$ 
matches $\bm{n}$ and $\bm{t}$ matches $\bm{y}$, the transformation of stiffness tensor can be written as: 
\begin{equation}
 C_{nrts} = a_{ni}a_{tj}a_{rk}a_{sl}C_{ijkl}.
\end{equation}

Similarly, the transformation of stress and strain tensor(including the eigenstrain tensor) can be written as:
\begin{align}
 \sigma_{nn} &= a_{nx}a_{nx}\sigma_{xx}, \quad \sigma_{nt} = a_{nx}a_{ty}\sigma_{xy}, \\ \nonumber
  \epsilon_{nn} &= a_{nx}a_{nx}\epsilon_{xx}, \quad \epsilon_{nt} = a_{nx}a_{ty}\epsilon_{xy}.
\end{align}
where, the transformation matrix can expressed using rotation matrix, where we 
elaborate each element of rotation matrix as follows:
\begin{align*}
 a_{xn} &= cos\left(tan^{-1}\dfrac{n_x}{n_y}\right) = a_{nx}, \\
 a_{xt} &= -sin\left(tan^{-1}\dfrac{n_x}{n_y}\right) = a_{tx}, \\
 a_{yn} &= sin\left(tan^{-1}\dfrac{n_x}{n_y}\right) = a_{ny}, \\
 a_{yt} &= -cos\left(tan^{-1}\dfrac{n_x}{n_y}\right) = a_{ty}. \\
\end{align*}

The elastic energy of each of the phases writes as, 
\begin{equation}
 f_{el} = \dfrac{1}{2}\sigma_{ij}\left(\epsilon_{ij}-\epsilon^{*}_{ij}\right),
\end{equation}
which can be further elaborated for transformed co-ordinate system, where we express elastic 
energy in terms of normal and tangential components of stresses and strains as, 
\begin{equation}
 f_{el} = \dfrac{1}{2}\left(\sigma_{nn}\left(\epsilon_{nn}-\epsilon^{*}_{nn}\right) + 2\sigma_{nt}\left(\epsilon_{nt}-\epsilon^{*}_{nt}\right) 
 + \sigma_{tt}\left(\epsilon_{tt}-\epsilon^{*}_{tt}\right)\right).
 \label{elast_equation_tensorial}
\end{equation}

The interpolated elastic energy can then be described as, 
\begin{align}
 f_{el}\left(\bm{u},\phi\right) &= f^{\alpha}_{el}(\epsilon^{\alpha}_{nn}\left(\bm{u},\phi\right),\epsilon^{\alpha}_{nt}\left(\bm{u},\phi\right),\epsilon_{tt})h(\phi) \nonumber\\
	  &+ f^{\beta}_{el}(\epsilon^{\beta}_{nn}\left(\bm{u},\phi\right),\epsilon^{\beta}_{nt}\left(\bm{u},\phi\right),\epsilon_{tt})(1 - h(\phi)).
	  \label{interpolated_elastic_energy}
\end{align}

Note, the usage of the superscripts on some of the terms, with the terms $\epsilon_{nn}, \epsilon_{nt}, \sigma_{tt}$, 
while the others $\epsilon_{tt},\sigma_{nn},\sigma_{nt}$ are left free, which is related to 
the jump conditions at the interface. In this interpolation scheme, the normal stress components $\sigma_{nn},\sigma_{nt}$
and $\epsilon_{tt}$ are continuous variables across the interface, which are derived from the condition of 
continuity of normal tractions at the interface (in the absence of interfacial stresses) 
and the Hadamard boundary conditions \cite{hadamard_silhavy2013} respectively, while the others $\epsilon_{nn},\epsilon_{nt},\sigma_{tt}$ 
have a discontinuity in the sharp interface free-boundary problem, which would translate to a 
smooth variation across the interface in the diffuse interface description. In order to affect this 
idea, the following scheme for the determination of the stress and strain components is adopted.

Firstly the individual normal stress components of each of the phases are written 
down explicitly as a function of the stiffness and the strains as,
\begin{align*}
 \sigma_{nn} &=  C^{\alpha,\beta}_{nntt}(\epsilon_{tt}-\epsilon^{*\alpha,\beta}_{tt}) 
    + C^{\alpha,\beta}_{nnnn}(\epsilon^{\alpha,\beta}_{nn} -\epsilon^{*\alpha,\beta}_{nn}) \\
    &+ 2C^{\alpha,\beta}_{nnnt}\left(\epsilon^{\alpha,\beta}_{nt}-\epsilon^{*\alpha,\beta}_{nt}\right) \\ 
  \sigma_{nt} &=  C^{\alpha,\beta}_{nttt}(\epsilon_{tt}-\epsilon^{*\alpha,\beta}_{tt}) 
    + C^{\alpha,\beta}_{ntnn}(\epsilon^{\alpha,\beta}_{nn} -\epsilon^{*\alpha,\beta}_{nn})  \\
    &+ 2C^{\alpha,\beta}_{ntnt}\left(\epsilon^{\alpha,\beta}_{nt}-\epsilon^{*\alpha,\beta}_{nt}\right),
\end{align*}
where, $\epsilon^*$ is the eigenstrain at phase interface produced due to lattice mismatch 
of precipitate and matrix. As mentioned before, we will be assuming that 
the eigenstrain is accommodated in the $\alpha$ phase, such that the 
eigenstrain in the $\beta$ phase is zero. After some re-arrangement, 
the individual non-homogeneous normal strain-fields in either phase can be 
written as functions of the continuous variables which are 
the normal stresses and the tangential strain, and written as, 
\begin{align}
 \label{strain_alpha}
   &\begin{bmatrix}
    (\epsilon^{\alpha}_{nn} - \epsilon^{*\alpha}_{nn}) \\
    (\epsilon^{\alpha}_{nt} - \epsilon^{*\alpha}_{nt})
   \end{bmatrix} \\ \nonumber
  = 
   &\begin{bmatrix}
    C^{\alpha}_{nnnn} & 2C^{\alpha}_{nnnt} \\ 
    C^{\alpha}_{ntnn} & 2C^{\alpha}_{ntnt} 
   \end{bmatrix}^{-1}
   \begin{bmatrix}
    \sigma_{nn} - C^{\alpha}_{nntt}\left(\epsilon_{tt} - \epsilon^{*\alpha}_{tt}\right) \\
    \sigma_{nt} - C^{\alpha}_{nttt}\left(\epsilon_{tt} - \epsilon^{*\alpha}_{tt}\right)
   \end{bmatrix}.
   \label{strain_alpha}
\end{align}
Similarly, for matrix phase, where there is no eigenstrain, the phase normal strains read, 
\begin{equation}
   \begin{bmatrix}
    \epsilon^{\beta}_{nn} \\
    \epsilon^{\beta}_{nt}
   \end{bmatrix}
  = 
   \begin{bmatrix}
    C^{\beta}_{nnnn} & 2C^{\beta}_{nnnt} \\ 
    C^{\beta}_{ntnn} & 2C^{\beta}_{ntnt} 
   \end{bmatrix}^{-1}
   \begin{bmatrix}
    \sigma_{nn} - C^{\beta}_{nntt}\epsilon_{tt} \\
    \sigma_{nt} - C^{\beta}_{nttt}\epsilon_{tt}    
   \end{bmatrix}.
   \label{strain_beta}
\end{equation}

In order to have a smooth variation of non-homogeneous variables, we 
impose the following interpolation upon the individual strain 
components as, 
\begin{equation}
   \begin{bmatrix}
    \epsilon_{nn} \\
    \epsilon_{nt}
   \end{bmatrix}
   =
      \begin{bmatrix}
    \epsilon^{\alpha}_{nn} \\
    \epsilon^{\alpha}_{nt}
   \end{bmatrix}
   h(\phi)
   + 
   \begin{bmatrix}
    \epsilon^{\beta}_{nn} \\
    \epsilon^{\beta}_{nt}
   \end{bmatrix}
   (1 - h(\phi)).
   \label{strain_inter}
\end{equation}

Expressing the inverse matrices in the previous relations as 
$S^{\alpha}_{nt}$ and $S^{\beta}_{nt}$ for 
the respective phases and further simplifying for the stress 
tensor $\sigma_{nn}$ and $\sigma_{nt}$ we can derive, 
\begin{align}
   \begin{bmatrix}
    \sigma_{nn}  \\
    \sigma_{nt}     
   \end{bmatrix}
   =
   &\left(S^{\alpha}_{nt}h(\phi) + S^{\beta}_{nt}(1 - h(\phi)) \right) \\
   &\Big\lbrace
      \begin{bmatrix}
    \epsilon_{nn} - \epsilon^{*\alpha}_{nn}h\left(\phi\right) \\
    \epsilon_{nt} - \epsilon^{*\alpha}_{nt}h\left(\phi\right) 
   \end{bmatrix}
   + S^{\alpha}_{nt}
    \begin{bmatrix}
     C^{\alpha}_{nntt}(\epsilon_{tt} - \epsilon^{*\alpha}_{tt}) \\ \nonumber
     C^{\alpha}_{nttt}(\epsilon_{tt} - \epsilon^{*\alpha}_{tt})    
    \end{bmatrix}
    h(\phi) \\
   &+ S^{\beta}_{nt}
    \begin{bmatrix}
     C^{\beta}_{nntt}\epsilon_{tt} \\ \nonumber
     C^{\beta}_{nttt}\epsilon_{tt}    
    \end{bmatrix}
    (1 - h(\phi)) 
   \Big\rbrace,
\end{align}

where, the local strains $\epsilon_{nn},\epsilon_{tt},\epsilon_{nt}$ 
can be derived from the displacement using Eqn.\ref{strain} followed
by a co-ordinate transformation to the $\bm{n,t}$ space.
Thus, we can derive the values of $\epsilon^{\alpha,\beta}_{nn}$ and $\epsilon^{\alpha,\beta}_{nt}$ 
by inserting above relation in previous strain calculations, i.e 
Eqns.\ref{strain_alpha},\ref{strain_beta}.

The remaining stress component $\sigma_{tt}$ i.e. the tangential component of stress 
is also non-homogeneous and can be derived by firstly imposing a smooth interpolation
across the diffuse interface, i.e.
\begin{equation}
 \sigma_{tt} = \sigma^{\alpha}_{tt}h(\phi) + \sigma^{\beta}_{tt}(1 - h(\phi)),
\end{equation}

where each of the phase stress components $\sigma^{\alpha,\beta}_{tt}$ can be written 
as a function of the normal strain components that have already been derived and the 
continuous tangential strain component that is just a function of the local displacement.

Therefore, relations for $\sigma^{\alpha}_{tt}$ and $\sigma^{\beta}_{tt}$ write as, 
\begin{align}
 \sigma^{\alpha}_{tt} &= C^{\alpha}_{ttnn}(\epsilon^{\alpha}_{nn} - \epsilon^{*\alpha}_{nn}) + 2C^{\alpha}_{ttnt}\left(\epsilon^{\alpha}_{nt}- \epsilon^{*\alpha}_{nt}\right)  \\ \nonumber
			      &+ C^{\alpha}_{tttt}(\epsilon^{\alpha}_{tt} - \epsilon^{*\alpha}_{tt}), \\
 \sigma^{\beta}_{tt} &= C^{\beta}_{ttnn}\epsilon^{\beta}_{nn} + 2C^{\beta}_{ttnt}\epsilon^{\beta}_{nt} 
			      + C^{\beta}_{tttt}\epsilon^{\beta}_{tt}.
\end{align}

Since, the values of $\epsilon^{\alpha,\beta}_{nn}$, $\epsilon^{\alpha,\beta}_{nt}$ are 
already determined in terms of $\epsilon_{nn}$, $\epsilon_{nt}$ and $\epsilon_{tt}$, the 
value of $\sigma^{\alpha,\beta}_{tt}$ can be solved using preceding relations. Inserting 
them into equation for $\sigma_{tt}$, helps to determine the final stress component as 
function of strains. 

In this event, the variational derivative of elastic energy of system at constant stress, 
strain and displacement can be derived from Eqn.\ref{interpolated_elastic_energy} as, 
\begin{align}
  \left(\dfrac{\partial f_{el}}{\partial\phi}\right)_{\sigma,\epsilon,u} 
  = &\left(f^{\alpha}_{el}-f^{\beta}_{el}\right)\dfrac{\partial h(\phi)}{\partial\phi} 
   + \sum_{\alpha}\dfrac{\partial f^{\alpha}_{el}}{\partial \epsilon^{\alpha}_{nn}} \dfrac{\partial 
   \epsilon^{\alpha}_{nn}}{\partial\phi}h_\alpha(\phi) \\	\nonumber
  &+ 2\sum_{\alpha}\dfrac{\partial f^{\alpha}_{el}}{\partial \epsilon^{\alpha}_{nt}} \dfrac{\partial 
   \epsilon^{\alpha}_{nt}}{\partial\phi}h_\alpha(\phi),
\end{align}
where for conciseness we introduce the notation $h_\alpha\left(\phi\right)=h\left(\phi\right)$
and $h_\beta\left(\phi\right)=1-h\left(\phi\right)$, and summations on the R.H.S, run 
over the phases, $\alpha,\beta$. The derivative with respect to the variable $\epsilon_{tt}$ returns zero, 
as it is a homogeneous quantity. In order to determine the derivative of 
non-homogeneous variables we utilize Eqn.~\ref{strain_inter}. Differentiating 
Eqn.~\ref{strain_inter} at constant displacement and strain, gives us following relations, 
\begin{align}
  -\begin{bmatrix}
    \epsilon^{\alpha}_{nn} - \epsilon^{\beta}_{nn} \\
    \epsilon^{\alpha}_{nt} - \epsilon^{\beta}_{nt}
   \end{bmatrix} \dfrac{\partial h_{\alpha}(\phi)}{\partial \phi}
  = 
   \sum_{\alpha}   \begin{bmatrix}
		   \dfrac{\partial\epsilon^{\alpha}_{nn}}{\partial\phi} \\
		   \dfrac{\partial\epsilon^{\alpha}_{nt}}{\partial\phi}
		   \end{bmatrix} h_{\alpha}(\phi).
\end{align}

Also recall that,
\begin{align}
 \dfrac{\partial f^{\alpha,\beta}_{el}}{\partial \epsilon^{\alpha,\beta}_{nn}} = \sigma_{nn}, \quad
 \dfrac{\partial f^{\alpha,\beta}_{el}}{\partial \epsilon^{\alpha,\beta}_{nt}} = \sigma_{nt},
\end{align}
which follow from the definition of the elastic energy.

Substituting these values and rewriting the variational derivative, we derive,
\begin{align}
 \left(\dfrac{\partial f_{el}}{\partial\phi}\right)
 &= \left(f^{\alpha}_{el}-f^{\beta}_{el}-\sigma_{nn}(\epsilon^{\alpha}_{nn} -\epsilon^{\beta}_{nn})\right)
					\dfrac{\partial h\left(\phi\right)}{\partial \phi} \\ \nonumber
                                       &-2\sigma_{nt}(\epsilon^{\alpha}_{nt}-\epsilon^{\beta}_{nt})
   \dfrac{\partial h\left(\phi\right)}{\partial \phi}.
\end{align}

This relation can be simplified further by substituting
the values for $f^{\alpha}_{el}$ and $f^{\beta}_{el}$ explicitly
in terms of the stresses and the strains as, 
\begin{align}
\left(\dfrac{\partial f_{el}\left(\bm{u},\phi\right)}{\partial\phi}\right) &= 
 \Big[-\dfrac{1}{2}\sigma_{nn}\left(\epsilon^{\alpha}_{nn}-\epsilon^{\beta}_{nn}\right) - \sigma_{nt}\left(\epsilon^{\alpha}_{nt} - \epsilon^{\beta}_{nt}\right) \nonumber\\
 &-\dfrac{1}{2}\sigma_{tt}^{\alpha}\epsilon^{*\alpha}_{tt}
 -\dfrac{1}{2}\sigma_{nn}\epsilon^{*\alpha}_{nn} - \sigma_{nt}\epsilon^{*\alpha}_{nt} \\ \nonumber
 &+ \dfrac{1}{2}\left(\sigma_{tt}^{\alpha} - \sigma_{tt}^{\beta}\right)\epsilon_{tt}\Big] \dfrac{\partial h\left(\phi\right)}{\partial \phi}.
\end{align}

\subsection{Conservation of volume}

The remaining part of the evolution equation in Eqn.\ref{phi_evolve} that is yet 
to be determined is the Lagrange parameter $\lambda_\beta$ which would conserve 
the volume during interface evolution. Volume conservation is affected through
the constraint, 

\begin{align}
  \int h\left(\phi\right) \cdot dx = const \quad {or}\\
  \int \delta h\left(\phi\right) dx = 0,
  \label{volume_constraint}
\end{align}

where $\delta h\left(\phi\right)$ is the change in the value of $h\left(\phi\right)$ at a given 
spatial location. Reformulation, of this condition in discrete terms is performed
in the following manner. From the Eqn.\ref{phi_evolve}, we have the rate of change
of the order parameter at a given location, i.e
\begin{align}
 \tau W \dfrac{\partial \phi}{\partial t} &= \textrm{rhs}_\alpha - \lambda_\beta h^{\prime}\left(\phi\right),
\end{align}
where the term $\textrm{rhs}_\alpha$ constitute all the terms in the evolution equation of 
the order parameter in Eqn.\ref{phi_evolve} leaving out the Lagrange parameter. In order
to affect the volume constraint as given by Eqn.\ref{volume_constraint}, 
the Lagrange parameter $\lambda_\beta$ is computed as, 

\begin{align}
 \lambda_{\beta} = \frac{\sum_{V}rhs_\alpha}{\sum_{V}h^{\prime}(\phi)}, 
\end{align}

where, the summation $\sum_V$ is over the entire volume. This essentially 
ensures that the summation of all the changes in the order parameter
over the entire volume returns zero, thus affecting the volume constraint
in the discrete framework.

\subsection{Mechanical equilibrium}

As a final aspect, what remains is the computation of the displacement 
fields as a function of the spatial distribution of the order parameter. 
This is done iteratively by solving the damped wave equation written 
as, 

\begin{align}
  \rho\frac{d^2\bm{u}}{dt^2} + b\frac{d\bm{u}}{dt} &= \nabla \cdot \bm{\sigma}, 
\end{align}

that is solved until the equilibrium is reached, i.e $\nabla\cdot\bm{\sigma}=\bm{0}$.
The terms $\rho$ and $b$ are chosen such that the convergence is achieved in the 
fastest possible time. The computation of the stresses differs depending upon the 
interpolation schemes used for estimating the elastic energies. For the case of 
the Khachaturyan interpolation, the stresses at every point can be readily computed
as a partial derivative of the elastic energy as $\sigma_{ij}=\dfrac{\partial f_{el}\left(\bm{u},\phi\right)}{\partial \epsilon_{ij}}$, 
while for the tensorial interpolation, the estimation of the stresses is done 
differently as laid out in the preceding section (see sec. \ref{Tensorial}).

The optimization procedure for finding the equilibrium shape is performed
by solving the evolution of the order parameter $\phi$ in Eqn.\ref{phi_evolve}
along with the equation of mechanical equilibrium at each time step, until a
converged shape is reached. 

\section{Results}
\subsection{Model parameters}

In this section, we list out the material parameters and the non-dimensionalization 
scheme that will be used in the subsequent sections. Firstly, we will limit ourselves
to isotropic and cubic systems in two dimensions, such that the stiffness tensor can be 
be simplified. These systems can be generically defined in the following manner,
where we use the commonly used short-hand notation for the non-zero 
stiffness components, $C_{11}=C_{1111}$, $C_{22}=C_{2222}$, 
$C_{12}=C_{1122}$ and $C_{44}=C_{1212}$, where additionally
$C_{11}=C_{22}$ because of symmetry considerations. Thereafter,
these terms can be derived in terms of the known material 
parameters, those are the Zener anisotropy ($Az$), 
Poison ratio ($\nu$) and shear modulus ($\mu$) and
can be written as, 
\begin{align}
  C_{44} = \mu, \quad
  C_{12} = 2\nu\left(\frac{C_{44}}{1-2\nu}\right), \quad
  C_{11} = C_{12} + \frac{2C_{44}}{A_z}.
\end{align}

The eigenstrain matrix will be considered diagonal in the 
Cartesian co-ordinate system and reads,
\begin{align}
  \epsilon^*
  =
  \begin{bmatrix}
   \epsilon^*_{xx} & 0 \\
   0 & \epsilon^*_{yy}
  \end{bmatrix}.
\end{align}

We use a non-dimensionalization scheme where the energy scale is set 
by the interfacial energy scale $1.0J/m^{2}$ divided by 
the scale of the shear modulus $1\times10^{9}J/m^{3}$ that yields 
a length scale $l^{*}=1$nm. In the paper, hence all the parameters
will be reported in the terms of non-dimensional units. Unless otherwise specified, 
all results are produced with $\mu_{mat} =125$, $\nu_{ppt}=\nu_{mat}=0.3$ and $A_z$ varies 
from 0.3 to 3.0. When $A_z=1.0$, elastic constants become isotropic. When $A_z$ is greater than 
unity, elastically soft directions are $<100>$, whereas elastically hard directions are $<110>$.
Similarly, in the case where $A_z$ is less than unity, elastically soft (hard) directions are 
$<110>$ ($<100>$). For all the cases, the precipitate and matrix have the same magnitude of $A_z$.

The diagonal components of misfit strain tensor are assumed to be aligned along $<100>$ directions 
in (001) plane of cubic system, whereas, off-diagonal terms are zero. The misfit strain or 
eigenstrain ($\epsilon^*$) can be dilatational i.e. same along principle directions 
($\epsilon^*_{xx}=\epsilon^*_{yy}$) or tetragonal i.e. different along principle directions 
($\epsilon^*_{xx}\ne\epsilon^*_{yy}$). For different cases magnitude of eigenstrain varies from 
0.5 to $2 \%$.

Here we have implemented two dimensional system of matrix embedded with precipitate having 
certain misfit due to transformation at interface. Domain boundaries follow periodic conditions. 
The ratio of precipitate size to matrix is maintained 0.08, such that there is negligible 
interaction of displacement field at the boundaries. This resembles the condition of infinitely 
large matrix  containing an isolated precipitate without any influence of external stress. 
Interfacial energy between matrix and precipitate is assumed to be isotropic until 
specified. The magnitude of interfacial energy is considered to be 0.15.

The model is generic enough for incorporating any combination of 
anisotropies in elastic energy which are possible in 
two dimensional space. In addition, different stiffness values in the phases 
can also be modeled. In our model formulation, we will use the ratio of the 
shear moduli $\delta$ in order to characterize the degree of inhomogeneity, 
wherein, a softer precipitate is derived by a value of $\delta$ that is less than unity, 
and vice-versa for the case of a harder precipitate.

\subsection{Isotropic elastic energy}

As the first case of study, we consider isotropic elastic moduli ($A_z=1.0$) with dilatational misfit 
at interface between precipitate and matrix ($\epsilon_{xx}^*=\epsilon_{yy}^*=0.01$). 
For this case, we consider the precipitate to be softer than matrix i.e. the inhomogeneity ratio, 
$\delta= 0.5$. 

\begin{figure}[htbp!]
 \centering
 \includegraphics [width=0.8\linewidth]{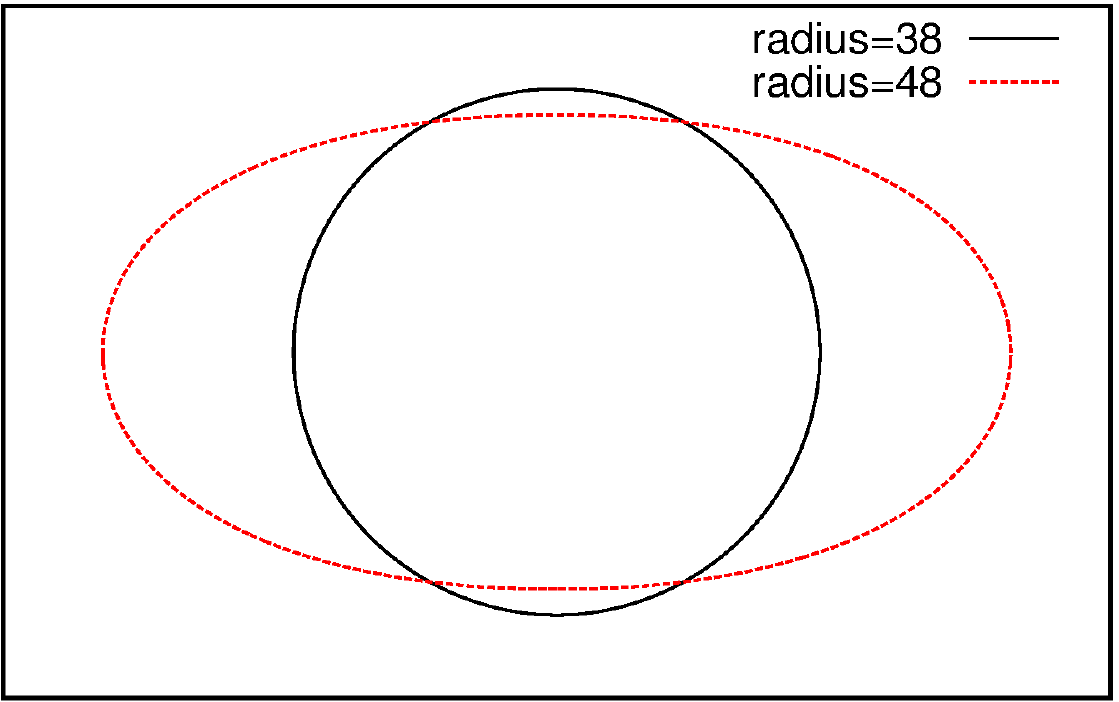} \\
 \vspace{0in}
 \caption{Equilibrium shapes of precipitate with radius 38 (thick line) 
	 and 48 (dotted line), with $\delta = 0.5$, $\mu_{matrix} = 125$, $\epsilon^* = 0.01$.}
 \label{eqm_shape_iso}
\end{figure}

We begin the simulation with an arbitrary shape as an initial state of precipitate e.g. an ellipse 
with a arbitrary aspect ratio, that is the ratio of the lengths of the major and 
the minor axes. We formulate a shape factor in terms of the major axis(c) and minor axis(a), 
which is expressed as $\rho=\frac{c-a}{c+a}$, that parameterizes the possible equilibrium 
shapes. An exemplary simulation showing the influence of the size is 
shown in Fig.~\ref{eqm_shape_iso}, where a precipitate with equivalent radius $R=38$ 
has circular shape. With increase in the size of precipitate, the precipitate transforms 
to an elliptical shape, which is captured in Fig.~\ref{eqm_shape_iso}, where a precipitate 
with larger size i.e. equivalent radius, $R=48$, acquires an ellipse-like configuration 
as the equilibrium shape.

This phenomenon has been theoretically studied by Johnson and Cahn~\cite{johnson1984elastically}, 
which they term as a symmetry breaking transition of a misfitting precipitate in elastically 
stretched matrix. The shape transition from a circular to an ellipse-like shape can be presented 
distinctly by plotting the shape factor as a function of 
the characteristic length. The characteristic length is 
defined as ratio of characteristic elastic energy to interfacial energy which can 
be written as follows,
\begin{equation}
 L = \frac{R\mu_{mat}\epsilon^{*2}}{\gamma},
\end{equation}
where, $R$ is equivalent radius of circular precipitate and $\gamma$ is the magnitude 
of surface energy. Here, we briefly describe the Johnson-Cahn theory for shape bifurcation
of a cylindrical precipitate.
The solution for an elastically induced shape bifurcation of an inclusion can be derived, 
only when precipitate is softer than matrix. To do so, we express 
surface energy ($f_s$) and elastic energy ($f_{el}$) in terms of the area of the precipitate 
(A) and shape factor ($\rho$).

Thus, the total surface energy can be expressed as a Taylor series expansion 
in terms of $\rho$,
\begin{equation}
 f_s = 2\sigma \sqrt{\pi A}\left(1 + \frac{3}{4}\rho^2 + \frac{33}{64}\rho^4 +...\right).
 \label{sf_eng}
\end{equation}

In the same way, we express total elastic energy as,
\begin{align}
 f_{el} &= \frac{2A\epsilon^{*2}\mu_{ppt}}{1+\delta-2\nu_{ppt}}\bigg\{1 
				     -\frac{\delta(1-\delta)(1+\kappa)}{(1+\kappa\delta)(1+\delta-2\nu_{ppt})}\rho^2 \\ \nonumber
				    &+\frac{\delta(1-\delta)^2(1+\kappa)}{(1+\kappa\delta)^2(1+\delta-2\nu_{ppt})^2}\rho^4
				    +...\bigg\},
 \label{elast_eng}
\end{align}
where, $\epsilon^*$ is misfit at interface, $\delta=\mu_{ppt}/\mu_{mat}$ and $\kappa=3-4\nu$.

The total energy ($F^T$) is the sum of $f_s$ and $f_{el}$. Shape bifurcation 
takes place when the precipitate acquires a critical area $A_c$ at which 
the energy landscape that is plotted as a function of the shape 
factor has distinct minima corresponding to the bifurcated shapes.
This is akin to the example of classical spinodal decomposition where 
the compositions of the phases bifurcate below a critical 
temperature. Therefore, the critical size and the bifurcated
shapes beyond it, can be derived using the same common tangent
construction as in spinodal decomposition, with the shape-factor
being used similarly as the composition. Given, the symmetry
of the isotropic system, this equilibrium condition simplifies to,
\begin{align}
 \frac{dF^T}{d\rho}=0. \\ \nonumber
 \label{bifur_pt}
\end{align}
In this way, we take derivatives of the coefficients of the Taylor series from 
Eqn.~\ref{sf_eng} and Eqn.~\ref{elast_eng} w.r.t. shape factor ($\rho$), and add the 
respective terms to solve for Eqn.~\ref{bifur_pt} as,
\begin{align}
   &\rho\Bigg\{ \overbrace{3\sigma \sqrt{\pi A_c}}^{\bm{df^{1}_s}} 
     \quad -\overbrace{\frac{4A\epsilon^{*2}\mu^*\delta(1-\delta)(1+\kappa)}{(1+\kappa\delta)(1+\delta-2\nu^*)^2}}^{\bm{df^{1}_{el}}} \Bigg\} \\ \nonumber
 + &\rho^3\Bigg\{ \underbrace{\frac{33}{8}\sigma \sqrt{\pi A_c}}_{\bm{df^{2}_s}} 
     \quad + \underbrace{\frac{8\delta(1-\delta)^2(1+\kappa)A_c\epsilon^{*2}\mu^*}{(1+\kappa\delta)^2(1+\delta-2\nu^*)^4}}_{\bm{df^{2}_{el}}}\Bigg\}=0.
      \label{bifur_pt1}
\end{align}

By rearranging the terms, we get a stable solution for $\rho$ as shown in 
subsequent relation,
\begin{equation}
   \rho = \pm\sqrt{\dfrac{df^{1}_{el} - df^{1}_s}{df^{2}_{s}+df^{2}_{el}}}.
 \label{rho_eq}
\end{equation} 

\begin{figure}[htbp!]
 \centering
 \includegraphics [width=0.8\linewidth]{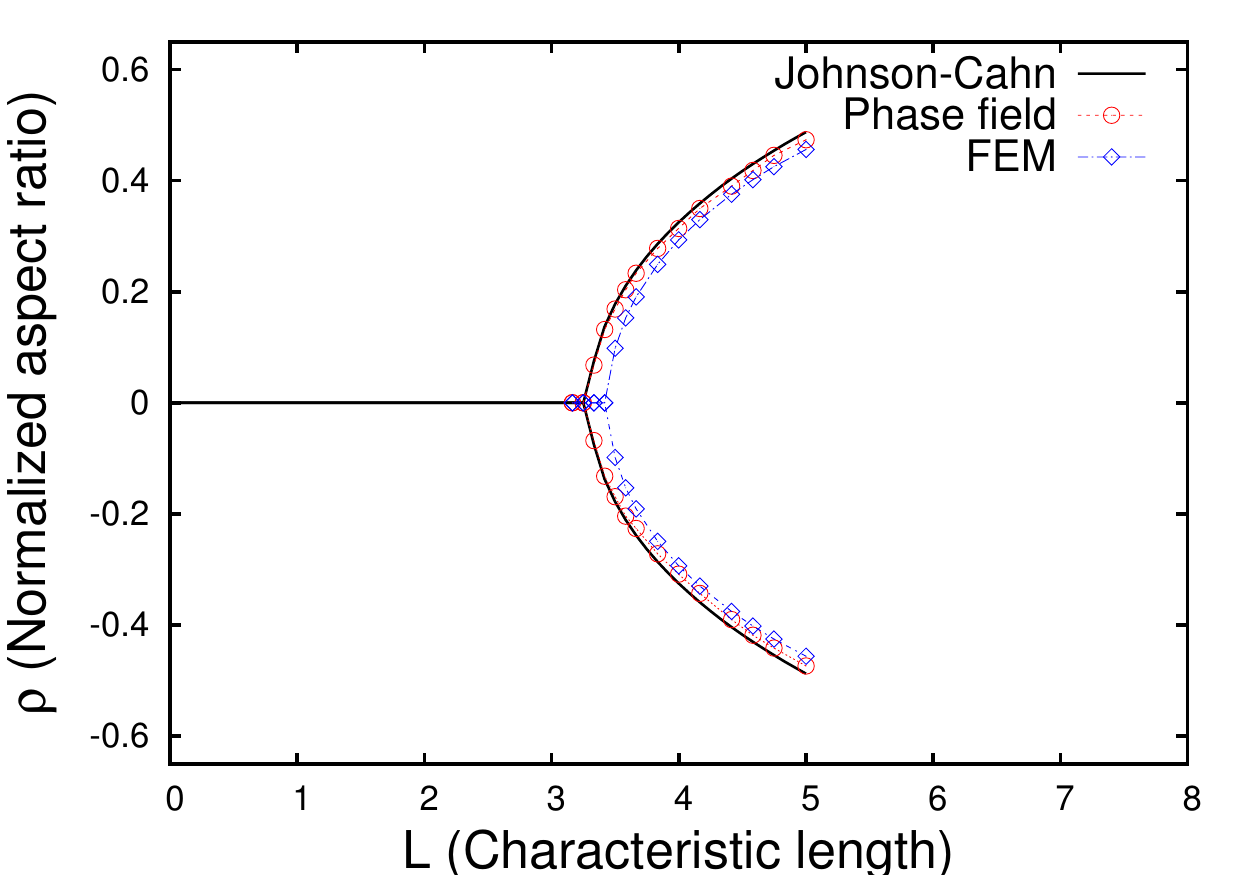} \\
 \vspace{0in}
 \caption{Plot depicts the shape bifurcation diagram- comparison of analytical solution 
  (thick black line) with phase field and FEM results, Az=1.0, $\epsilon^*=0.01$, $\delta=0.5$.}
 \label{bifurcation_plot_iso}
\end{figure}

By substituting for the variables in Eqn.~\ref{rho_eq}, we plot the shape factors corresponding to the 
equilibrium solution as a function of characteristic length. This is shown in Fig.~\ref{bifurcation_plot_iso}, where 
the thick dark line represents the analytical solution obtained from Eqn.~\ref{rho_eq}. 
Maxima or unstable solutions occur for $\rho=0$ for all the characteristic lengths beyond critical value. 

\begin{figure}[htbp!]
 \centering
 \includegraphics [width=0.8\linewidth]{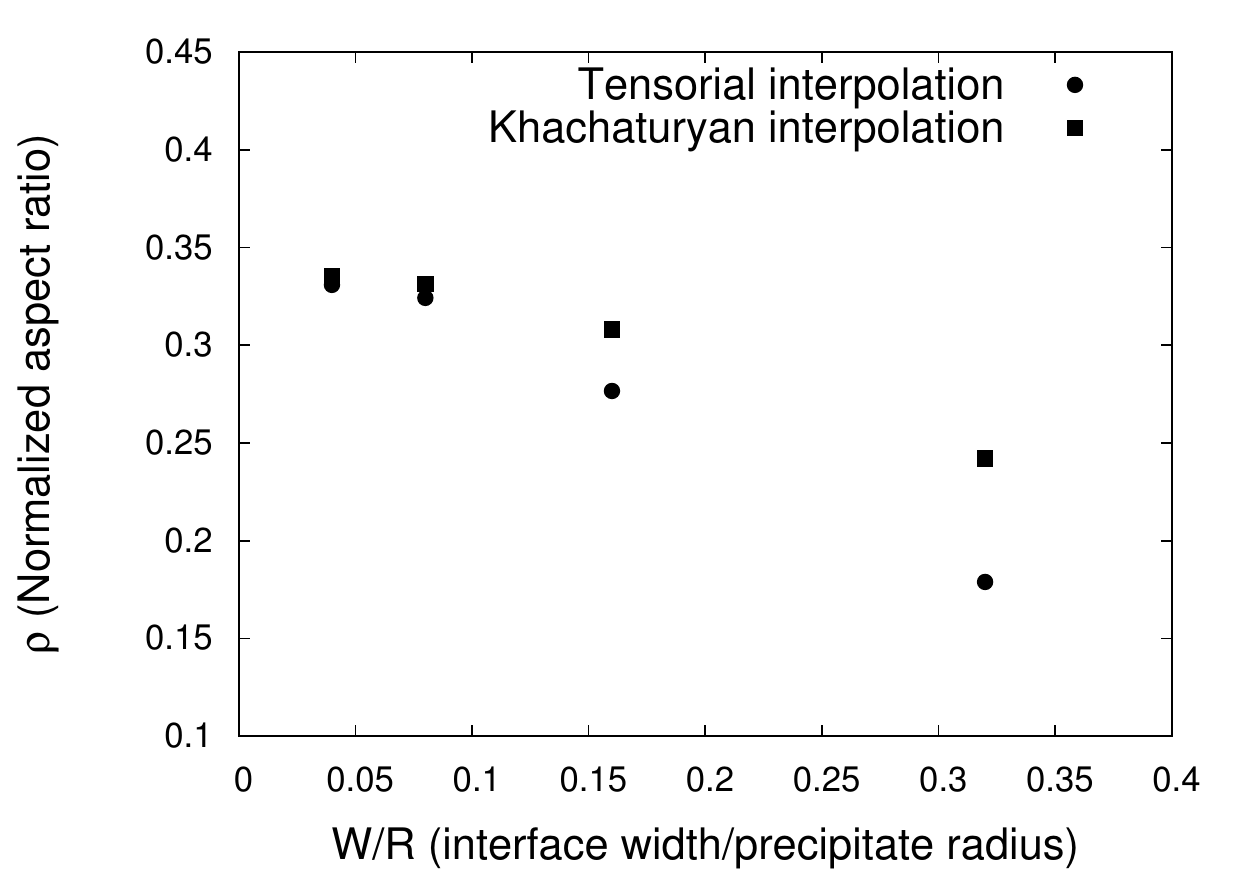} \\
 \vspace{0in}
 \caption{Plot shows the variation of normalized aspect ratio (shape factor) as function of $W/R$, 
			for $Az=1.0$, $\epsilon^*=0.01$, $\delta=0.5$.}
 \label{epsilon_convrg_iso}
\end{figure}

Before venturing into a critical comparison between analytical and the phase-field
results it is important to ensure the numerical accuracy with respect to the choice 
of the parameters particularly the choice of the interface width in the 
phase-field simulations. In the phase-field model that is described, the parameter that 
defines the diffuse-interface width $W$ is a degree of freedom which may be 
increased or decreased depending on the morphological and macroscopic length scales 
that are being modeled. This choice of the diffuse-interface width $W$ is 
typically chosen in a range such that the quantities being derived from the simulation
results remain invariant. So for example in the present scenario it would be
the shape-factor of the precipitate and its variation with the change in the 
interface widths allows us to determine the range of $W$ in which the
shape-factor is relatively constant.  We have performed this convergence test
for both the interpolation methods (recall from the model formulation: Khachaturyan and 
the Tensorial) described above and the comparison is described 
in Fig.~\ref{epsilon_convrg_iso}. 

We find that for both interpolation methods
relative invariance with change in the value of the interface width is achieved
for values of $W<=2$, while for values greater the variation is non-linear
and changes rapidly. Therefore, we have chosen values of $W=2$ for performing
the comparison between the different numerical methods (FEM and phase-field) and 
analytical calculations. It is noteworthy, that in general phase-field methods
show this variation with the change in the interface widths because of variation in the 
equilibrium phase-field profile arising out of a contribution to the total 
effective interfacial excesses from the bulk energetic terms that scales with 
the interface width. 
The other reason for variation with $W$ arises because of higher order corrections in the 
stress/strain profiles as a result of the imposition of an interface between the two 
bulk phases. Here, it is interesting that although the tensorial formulation is 
seemingly more correct with regards to the removal of the interfacial excess contribution 
\cite{schneider2015} to the interfacial 
energy, however, this leads to no advantage with respect to the choice of larger 
interfacial widths, in comparison to the Khachaturyan interpolation method. A possible
reason for this is the nature of the macroscopic length scale which in this case 
is the length over which the stress/strain profiles decay from the precipitate 
to the matrix, that typically are proportional to the local radius of curvature
of the precipitate. This implies that for smaller precipitates the decay is faster
and occurs over a shorter length and vice-versa for a larger precipitate. 
The accuracy of the phase-field method then, will naturally depend on the ratio of
the $W/R$ which is applicable for both interpolation methods.

Given, that the results of the variation of the shape-factor for different interface widths for
both interpolation methods, a seemingly qualitative conclusion that can be made
is that it is ratio of the interface width w.r.t to the macroscopic decay length 
of the stress/strain profiles and the error caused due to use of larger values leads to a more
stringent condition for the choice of the interface widths than the errors arising
from the contribution to the interfacial excesses due to incorrect interpolations
of the stresses/strains at the interface. Therefore, given that the Khachaturyan
method is numerically more efficient we have chosen this as the method in all
our future simulation studies. The convergence test w.r.t to the variations with 
the interface widths is repeated for the different bifurcation shapes both to 
verify and confirm the validity of the results as well as to perform the 
simulations in the most efficient manner by choosing the largest possible 
$W$ with the admissible deviation in shape-factor.

We now comment on the comparison between the different numerical and the 
analytical Johnson-Cahn theories with our phase-field results.
We have obtained the normalized aspect ratio ($\rho$) and plotted it as a function of normalized precipitate 
size (characteristic length), from phase field simulations. From the Fig.~\ref{bifurcation_plot_iso}, 
it is evident that precipitate has circular shape ($\rho=0$) below the critical point, and turns 
elliptical ($\rho\neq0$) beyond it. Also, transition from circle to ellipse-like shape is 
continuous, i.e. as $\rho$ approaches to zero, as $L$ approaches a critical value 
$L_c$. It is evident from Fig.~\ref{bifurcation_plot_iso} that, phase-field results are in 
very good agreement with the analytical solution derived from the work of 
Johnson and Cahn\cite{johnson1984elastically}, with a 
maximum error of 2.9\% in the studied range of characteristic lengths. It is expected that 
deviations occur for larger characteristic lengths where the truncation errors in the analytical 
expressions to approximate the surface and the elastic energies start to become larger. Therefore, 
the phase-field results should be more accurate here. 

Additionally, we compare our phase-field results with existing numerical methods,
such as the model adopted by Jog et.al.~\cite{jog2000symmetry}, 
where they used a finite element method coupled with a optimization technique to determine 
the equilibrium shape of coherent, misfitting precipitate in matrix, that is essentially a sharp 
interface technique. Using this method, we have reproduced the equilibrium shapes of precipitates, 
for the same set of conditions. It is clear that results 
obtained from the sharp interface model-FEM follows similar trends as that of the phase-field 
results (Fig.~\ref{bifurcation_plot_iso}). Note, that the errors are larger
near the critical point, which is expected to be better retrieved in the phase-field
method, given its greater resolution of the shape and lesser grid-anisotropy. Similarly
for larger precipitate shapes where the curvatures of the precipitates become larger at 
certain locations, again the phase-field method should yield 
a better estimate. Nevertheless, all three methods agree pretty well.

The critical size of the precipitate at shape transition can be determined from 
analytical solution as well as phase field simulations and FEM methods. 
Analytical equation gives $R_c=39.57(L_c=3.251)$, whereas the phase-field method yields 
$R_c=38.53(L_c=3.21)$ while the FEM yields $R_c=41.785(L_c=3.427)$. Thus, with these 
critical comparisons, we have benchmarked our phase field model quantitatively
with both the analytical solution and the sharp interface model. 

Moreover, the number of variants for bifurcated shape are infinite since all the directions 
in xy-plane are equivalent due to infinite fold rotation symmetry about the z-axis. 
We have confirmed this fact by starting the simulation with different orientations 
to the initial configurations, which equilibrate along different orientations but with same 
bulk energy. Also, for precipitates possessing rigid elastic moduli than that of matrix 
i.e. $\delta>1.0$, there is no bifurcation observed. This fact is also in agreement with 
the analytical solution, as there exits no real solution for cases where $\delta>1.0$. 

\subsection{Cubic anisotropy in elastic energy with dilatational misfit}

Anisotropy in the elastic energy arises from the variation of elastic constants in different 
directions. This deviation from the elastic isotropy is reflected in the increase 
in number of independent elastic constants. Here, we mainly consider cubic anisotropy in the elastic 
energy, as it is observed in several alloys while precipitation and growth.

\begin{figure}[htbp!]
 \centering
 \includegraphics [width=0.8\linewidth]{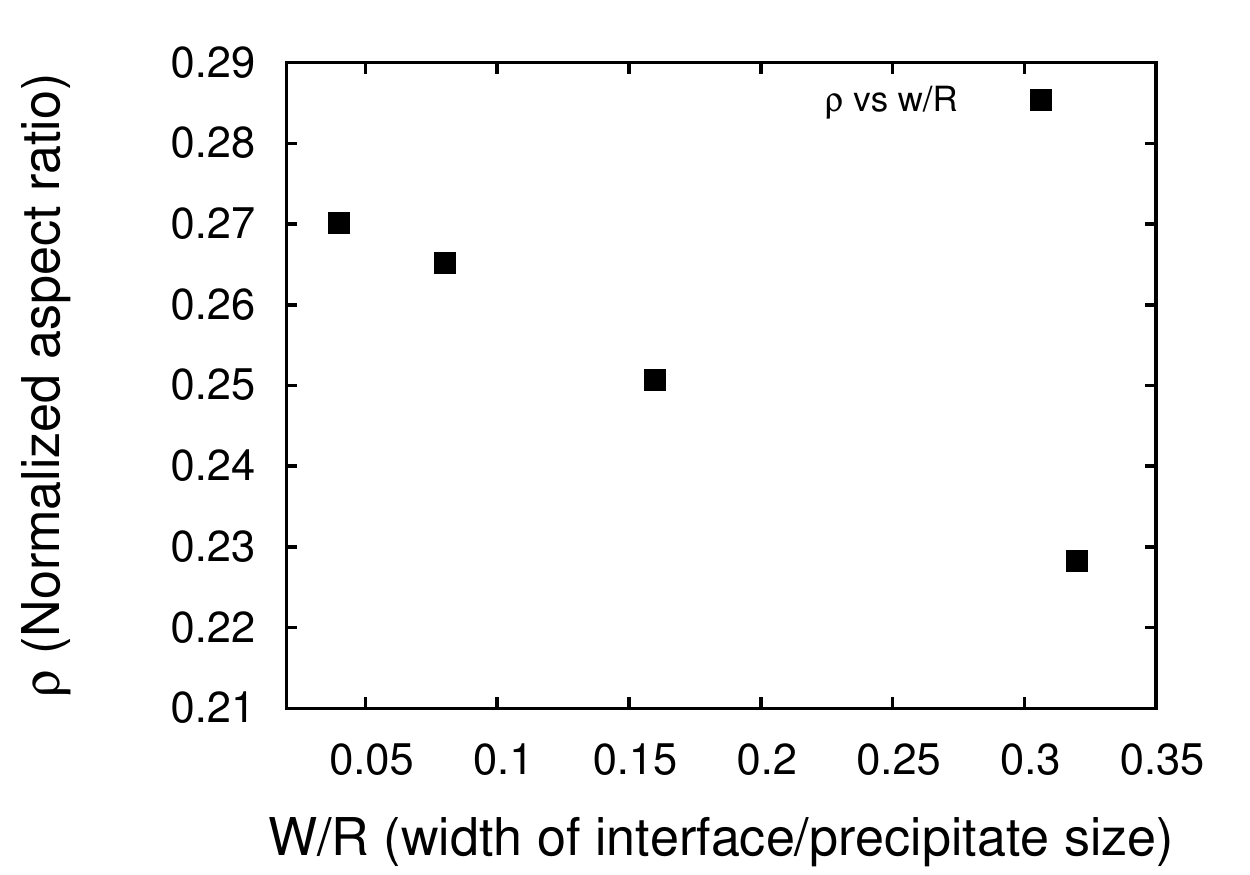} \\
 \vspace{0in}
 \caption{The variation of normalized aspect ratio as a function of $W/R$, for $Az=3.0$, $\epsilon^*=0.01$, $\delta=0.5$.}
 \label{epsilon_convrg_cubic}
\end{figure}

As explained in the previous section, eventually it is important to determine the range 
of $W$, where the shape factor remains constant for the different magnitudes of 
interfacial width. Anisotropy in the elastic elastic energy modifies the magnitude 
of elastic constants, thus changes stress/strain variation across the interface moving 
from the precipitate to matrix. As we have mentioned in the earlier section, we perform 
the $W$-convergence test, where the variations of shape-factor with 
the interface width is plotted in Fig.~\ref{epsilon_convrg_cubic}, 
There is not a significant variation in measured shape factor as 
a function of interfacial width i.e. in the given range of $W$, where the 
variation is weakly linear. But, as described in the earlier section, 
we choose an optimum value of $W=4$ with which we can efficiently run the simulations with 
an acceptable deviation in the calculated shape factor (i.e, an error of about 9\% from the 
value obtained by extrapolating to the y-axes or the case of $W=0$, that would effectively 
correspond to the sharp-interface limit). 

The model illustrated here can facilitate the variation of magnitude of different variables 
such as eigenstrain ($\epsilon^*$), anisotropy parameter ($A_z$), inhomogeneity ratio ($\delta$), 
elastic moduli, interfacial energy and precipitate size (R). The results we have presented 
here, show the combined influence of the above mentioned parameters on the equilibrium morphologies 
of the precipitate. Analogous to the previous section, simulations begin with a shape 
i.e. an ellipse with an arbitrary aspect ratio for the prescribed equivalent radius of precipitate(R), 
which eventually converge to an equilibrium shape. 

\begin{figure}[htbp!]
 \centering
 \includegraphics [width=0.8\linewidth]{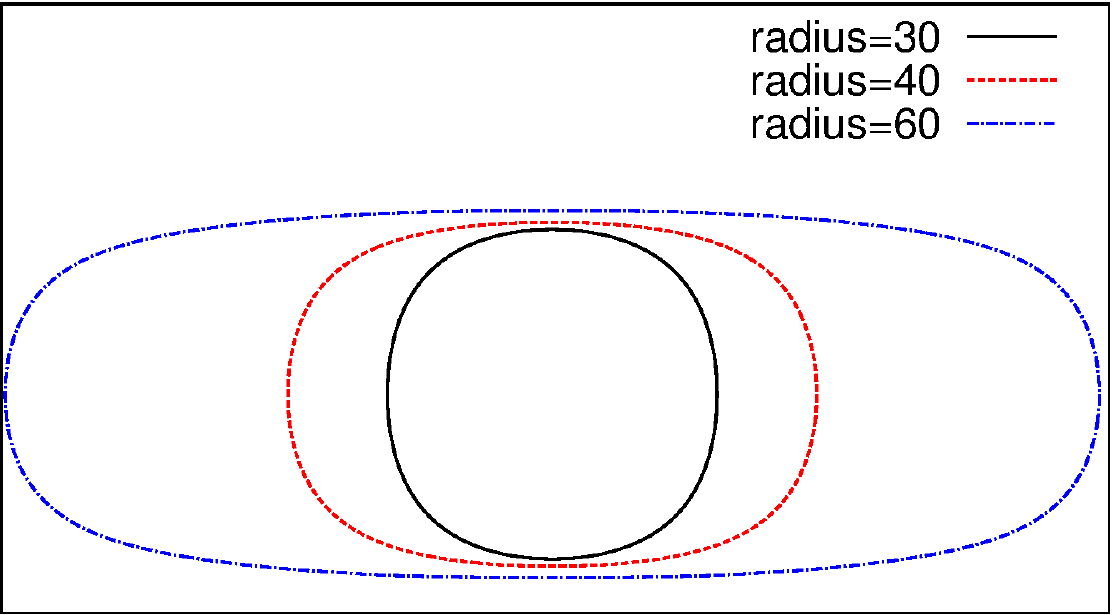} \\
 \vspace{0in}
 \caption{Equilibrium shapes of precipitate with cubic anisotropy 
          in elastic energy and with different sizes ($A_z=3.0$, $\epsilon^*=0.01$, $\delta=0.5$).}
 \label{eq_shapes_cubic}
\end{figure}

Here, the elastic moduli possess cubic anisotropy while the misfit is dilatational. Thus, we have used 
$A_z=3.0$, $\epsilon^*=0.01$, $\delta=0.5$, $\mu_{mat}=125$. With these initial conditions, 
the phase field simulations yield an equilibrium precipitate morphologies as a function of the characteristic 
length, which are illustrated in Fig.~\ref{eq_shapes_cubic}. Here, a precipitate with $R=30$, 
acquires cubic (square-like) shape with rounded corners as an effect of cubic anisotropy in 
the elastic energy. The precipitate faces are normal to $<100>$ directions, which are the elastically 
soft directions. In contrast to this, the precipitates with radii equal 
to 40 and 60 possess rectangular morphologies with rounded corners and elongated 
faces along one of the $<100>$ directions. Depending upon the orientation of the initial configuration 
of the precipitate i.e. the elliptical shape for a given equivalent radius, it converges to two variants
those are rectangle-like shapes, one aligned vertically (along $<010>$ direction) whereas, other along 
the horizontal axis (along $<100>$ direction).

\begin{figure}[htbp!]
 \centering
 \includegraphics [width=0.6\linewidth]{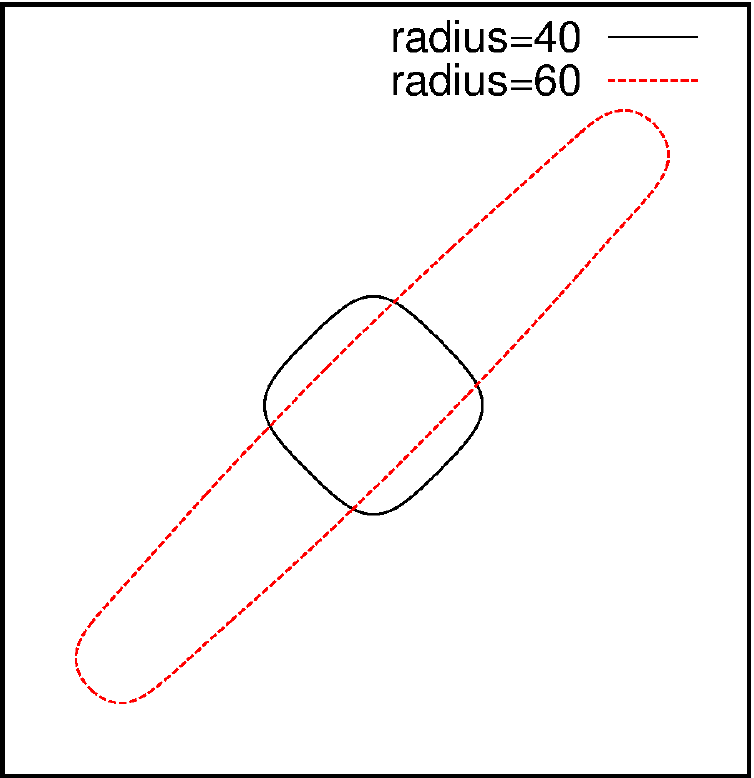} \\
 \vspace{0in}
 \caption{Equilibrium shapes of precipitate with cubic anisotropy 
          in elastic energy ($A_z=0.3$, $\epsilon^*=0.01$, $\delta=0.5$).}
 \label{eq_shapes_cubic2}
\end{figure}

Upon change in the anisotropy to a value of lesser unity, i.e. $A_z$ 
to 0.3, the equilibrium precipitate acquires a diamond like shape for R=40, which is 
shown in Fig.~\ref{eq_shapes_cubic2}. With $A_z<1$, the elastically softer directions now switch 
to $<110>$ directions. This is evident from the Fig.~\ref{eq_shapes_cubic2}, as the precipitate 
faces are oriented along $<110>$ directions. Further, with increasing the equivalent radius of 
precipitate, the equilibrium shape looses its four fold symmetry. The precipitate with increasing 
size tends to elongate along one of the elastically softer direction i.e. $<110>$ directions. 
This is captured by giving a slight orientation to the staring configuration, where the precipitate
eventually takes up a rectangle like morphology (oriented along $<110>$ direction), as shown in 
Fig.~\ref{eq_shapes_cubic2}. 

\begin{figure}[htbp!]
 \centering
 \includegraphics [width=0.9\linewidth]{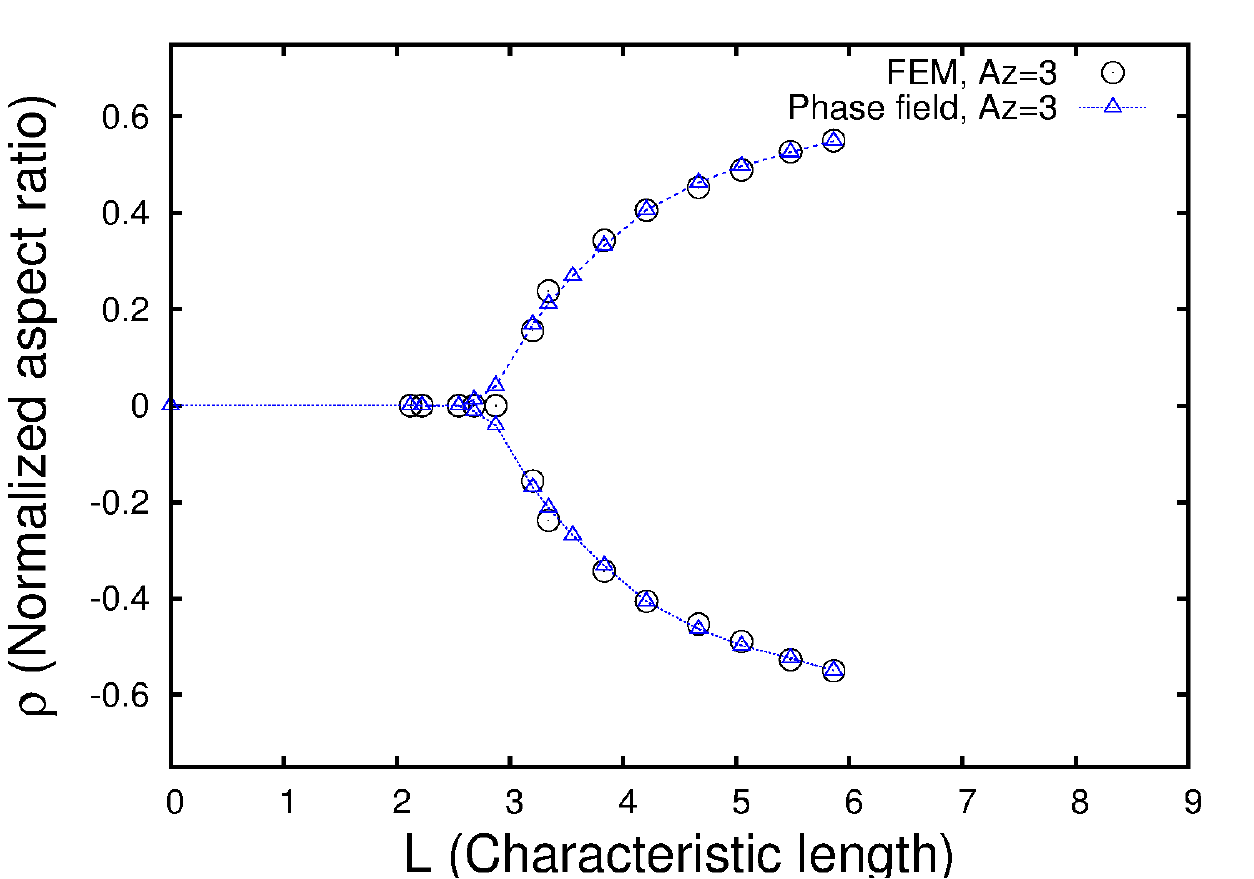} \\
 \vspace{0in}
 \caption{The shape bifurcation diagram with cubic anisotropy in 
          elastic energy ($A_z=3.0$, $\epsilon^*=0.01$, $\delta=0.5$), where the variation of aspect ratio is 
          plotted as function of characteristic length, a comparison of 
          phase field with FEM results.}
 \label{bifurcation_cubic}
\end{figure}

The influence of precipitate size on the equilibrium morphologies of the precipitate can be quantified 
by plotting the normalized aspect ratio (shape factor, $\rho$) as a function of characteristic 
length (L) of the precipitate, as a bifurcation diagram. 
We evaluate such a bifurcation diagram for the case of $A_z=3$. 
Here, the shape factor is calculated as the ratio of precipitate 
size measured along the horizontal axis to the vertical axis i.e. the size of precipitate along the 
elastically softer directions. Fig.~\ref{bifurcation_cubic} shows such a variation of the shape 
factor with respect to the precipitate size which reveals  that the critical size for the bifurcation from 
cubic $(\rho=0)$ to rectangle $(\rho\neq0)$, occurs for a value of $L=2.71$.

As illustrated in the previous section, we have obtained the results for the equilibrium morphologies of 
the precipitate with cubic anisotropy in the elastic energy using a sharp interface model (FEM), where 
the shape factor is calculated as a function of the characteristic length of the precipitate. This is shown 
in Fig.~\ref{bifurcation_cubic}, where the bifurcation diagram obtained from both the techniques i.e. 
phase-field as well as FEM are plotted against each other. It is evident that the bifurcation 
curves obtained from both simulation techniques, agree well with each other. 
Both techniques predict the critical characteristic length ($L_c$), that are close to 
each other i.e. $L_c$ retrieved from the phase field simulation equals to 2.71, whereas the one 
obtained from FEM equals to 2.87. Far away from the bifurcation point, the normalized aspect ratios 
obtained from phase field and FEM simulations predict nearly the same value, 
while near the bifurcation point ($L_c$) there is small 
variation, which is again expected as close to the critical point the 
resolution of the phase-field method should be better. 
In addition, the agreement between the two methods is also a critical additional benchmark
of the phase-field model in the absence of an analytical solution predicting the
shape factors for the case of cubic anisotropy. 

\subsection{Cubic anisotropy in elastic energy with tetragonal misfit}
\subsubsection{Misfit components with same sign}
In this subsection, we study the case of tetragonal misfit where the magnitude of 
eigenstrain is along the different principle directions but of same sign i.e. 
\begin{align}
  \epsilon^*
  =
  \begin{bmatrix}
   0.01 & 0 \\
   0 & 0.005
  \end{bmatrix},
\end{align}
where, we denote the degree of tetragonality with the parameter, $t=\epsilon^{*}_{xx}/\epsilon^{*}_{yy}=2$. 
We consider three different cases, 
where the magnitude of $A_z$ varies from 0.3 to 3.0 i.e. $Az<1$, $Az=1$ and $Az>1$. Similar to 
the previous cases, here the initial configuration of the precipitate is considered as an 
ellipse with an arbitrary aspect ratio and orientation. 

\begin{figure}[htbp!]
 \centering
 \includegraphics [width=0.3\linewidth]{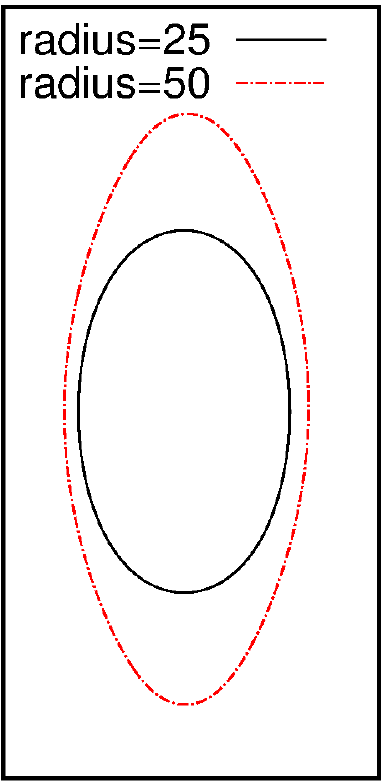} \\
 \vspace{0in}
 \caption{Equilibrium shapes of precipitate with tetragonal 
          misfit and different sizes, Az=3.0, t=+2.0, $\delta=1.0$.}
 \label{tetragonal_eq_shape1}
\end{figure}

Fig.~\ref{tetragonal_eq_shape1}, shows the equilibrium shapes of precipitate obtained with 
$A_z=3.0$, for two different equivalent radii of precipitate $(R=25, R=50)$. Here, 
the precipitate and matrix both possess the same elastic moduli i.e. $\delta=1.0$. It is clear 
that, the precipitate takes ellipse like shape for the given sizes. The precipitate 
elongates along y-direction, i.e. the direction along which the eigenstrain is lower. 
This implies that, with increase in the precipitate size, it will produce an equilibrium shape 
which aligns itself along the direction of smaller misfit while elongating along the 
same direction. Thus, for this situation, there is no 
shape bifurcation observed. Even for situations, where the magnitude of elastic anisotropy 
is $Az\geq1$ and $\delta\leq1$, the precipitate morphologies remain ellipse like for the larger equivalent radii 
of precipitates. 

\begin{figure}[htbp!]
 \centering
 \includegraphics [width=0.3\linewidth]{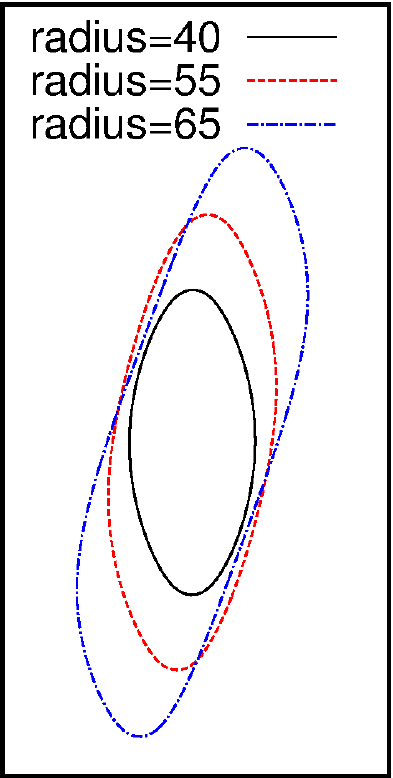} \\
 \vspace{0in}
 \caption{Equilibrium shapes of precipitate with tetragonal 
          misfit and with varying precipitate sizes, below (R=40) and above (R=55 and 65) bifurcation point, 
          Az=0.3, t=+2.0, $\delta=1.0$.}
  \label{tetragonal_eq_shape2}
\end{figure}

In the succeeding condition, we consider $A_z=0.3$ and $\delta=1.0$ with the same tetragonality. 
The equilibrium morphologies for the smaller sizes are only moderately different than those 
observed in the previous case. 
Fig.~\ref{tetragonal_eq_shape2} illustrates the results with $A_z=0.3$, where the precipitate 
size ranges from $R=40$ to $R=65$. The precipitate with comparatively 
smaller size takes up an ellipse like morphology, which is elongated along the direction of least 
misfit. However, with increase in the size of precipitate, its morphology changes from an ellipse 
like shape to a twisted diamond like shape as shown in the Fig.~\ref{tetragonal_eq_shape2}. 
Here, the precipitates with equivalent radii of $R=55$ and $R=65$ have lost their mirror symmetry 
that is observed for the smaller sizes.

This shape bifurcation can be understood as a competition between the tendency of
the precipitate to align along the elastically soft direction that is $<110>$ corresponding
to the choice of $A_z=0.3$ and the tetragonality influencing the shape towards
an elongated ellipse in the y-direction. The precipitate with smaller sizes tend to align along 
the lower misfit directions, while beyond the critical point the bifurcated shapes 
reflect the combined influence of both the elastic anisotropy and the tetragonality 
resulting in a twisted diamond shape with an orientation in between the 
lower misfit direction $(<100>)$ and elastically soft direction $(<110>)$. 

\begin{figure}[htbp!]
 \centering
 \includegraphics [width=0.3\linewidth]{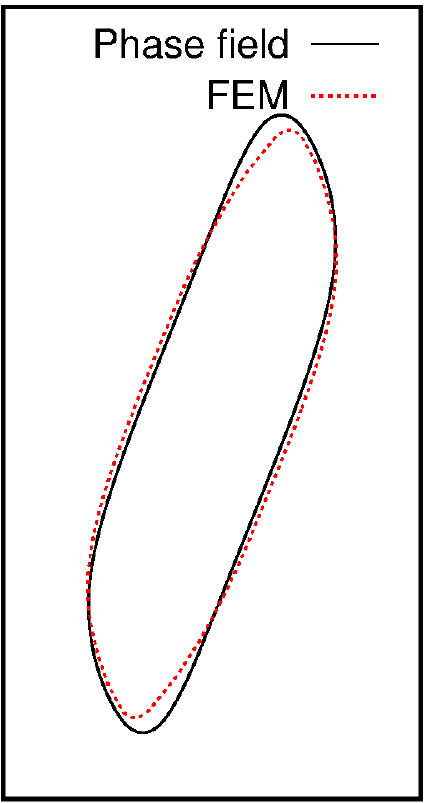} \\
 \vspace{0in}
 \caption{The comparison between equilibrium shapes of precipitate with equal area, 
			obtained from the phase field and FEM results, Az=0.3, t=+2.0, $\delta=1.0$.}
 \label{eq_shape_tetragonal_pf_vs_fem}
\end{figure}

Here, we will follow the same procedure as initially adopted in the previous sections. 
We compare the morphologies of the precipitate obtained from the phase field simulations with 
FEM, which is illustrated in Fig.~\ref{eq_shape_tetragonal_pf_vs_fem} for a given  
condition where, ($R=75$), $A_z=0.3$, $\delta=1.0$ and $t=2$. Again, we find an 
excellent agreement between the shapes computed from both numerical methods. 
In both the cases, volume occupied by the precipitates is same as well as 
both the precipitates are inclined similarly. Differences occur along the 
extended directions, where the curvatures from the FEM simulated shapes are 
slightly smaller compared to the ones produced from the phase-field simulations, 
which again given the increased spatial resolution of the phase-field 
method is not surprising.

In order to derive the bifurcation diagram, we have calculated the shape factor ($\rho$) as 
a function of the precipitate size i.e. characteristic length ($L$). The shape factor in this 
situation can be defined as:
\begin{equation}
 \rho = \sum_{i=1}^N \frac{X_i Y_i}{NV},
 \label{shape_factor_tetra}
\end{equation}
where, $X_i,Y_i$ are the co-ordinates on the precipitate-matrix interface considering 
center of mass of the precipitate is at origin, N is total the number of interfacial points and V is 
the volume of precipitate. This definition of the shape factor ensures that, it is equal to 
zero when the precipitate has mirror symmetry along the direction of least misfit i.e. the precipitate 
morphology is ellipse-like or elongated diamond, whereas it has non-zero values for a twisted 
diamond like shapes of the precipitates. 

\begin{figure}[htbp!]
 \centering
 \includegraphics [width=0.8\linewidth]{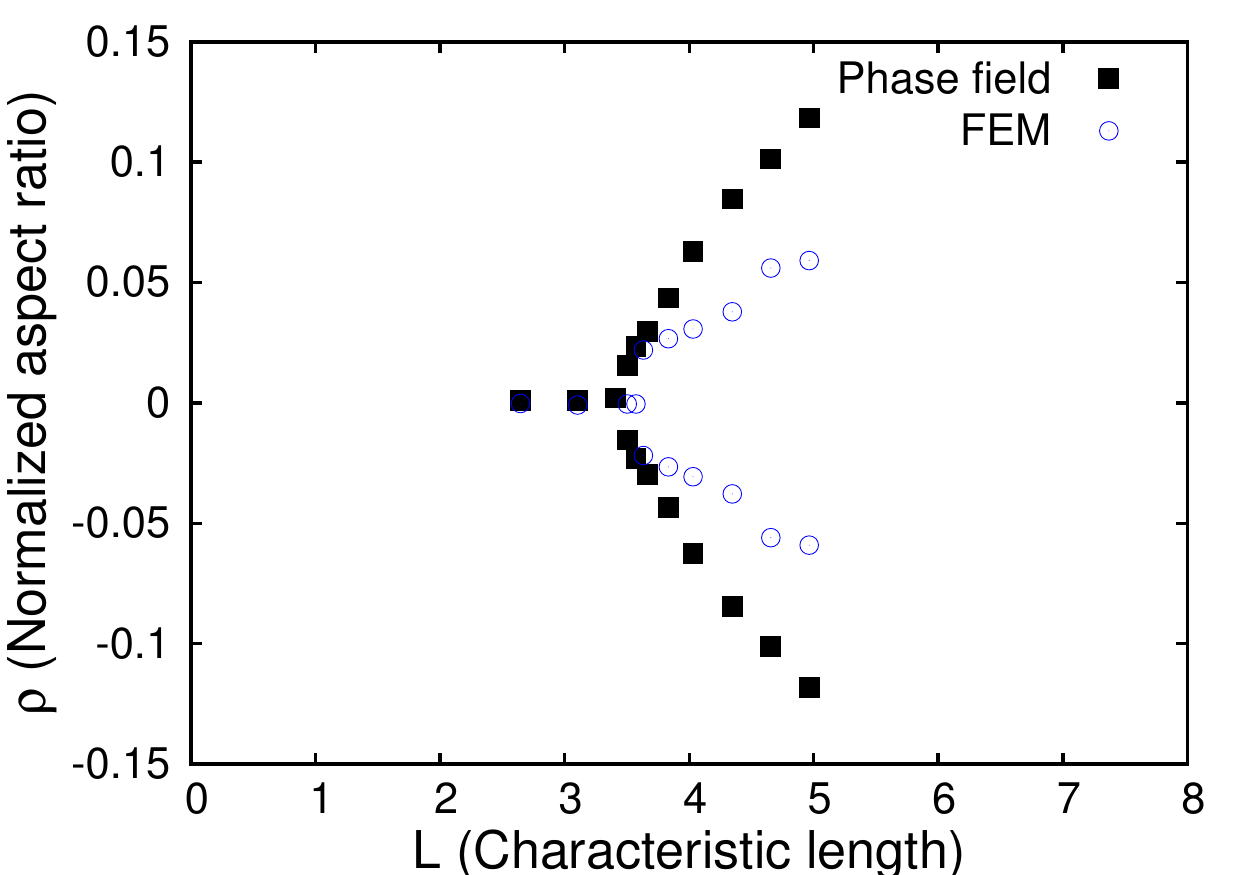} \\
 \vspace{0in}
 \caption{Shape bifurcation diagram for tetragonal misfit at t=2.0, 
	  comparison between phase field and FEM results ($A_z=0.3$, $\delta=1.0$), where $\rho$ is plotted 
	  as a function of characteristic length.}
 \label{bifurcation_tragonal}
\end{figure}

Fig.~\ref{bifurcation_tragonal}, depicts the calculated bifurcation diagram, which contains 
the results obtained from both the phase field as well as FEM simulations. 
We get a continuous transition of the shape 
factor beyond the bifurcation point from the phase field results, whereas the FEM results also 
give continuous transition but with a small jump at the bifurcation point and beyond. Phase field results 
show that the shape bifurcation occurs at a characteristic length of 3.41, whereas 
FEM simulations predict 3.49. It is observed that, beyond the bifurcation point the shape factors 
retrieved from the phase-field computations deviate from that of the FEM predictions. 
This difference in the calculations, might be due to the increased complexity in the 
shape which is possibly better resolved by the phase-field method.

\begin{figure}[htbp!]
 \centering
 \includegraphics [width=0.8\linewidth]{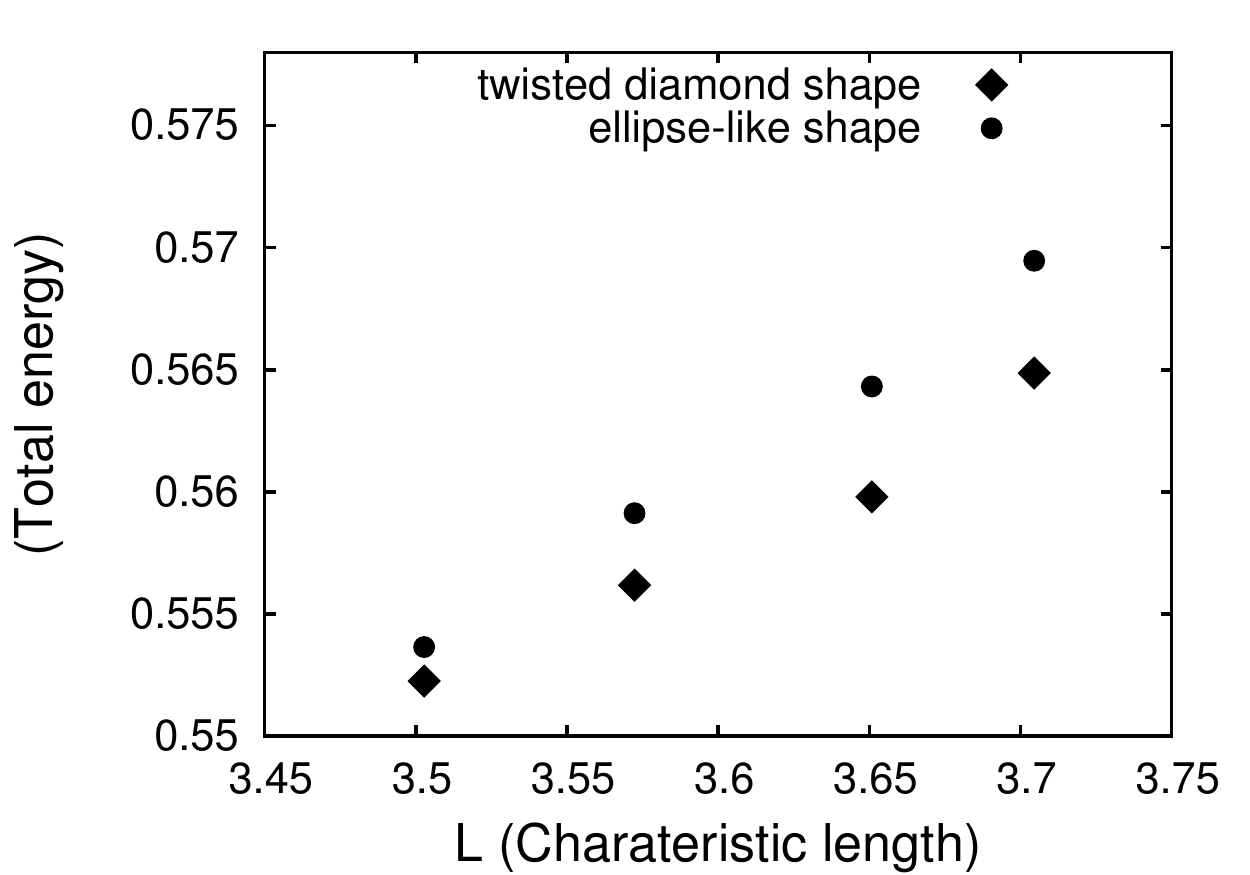} \\
 \vspace{0in}
 \caption{Variation of the elastic energy of precipitate as a function of characteristic length 
          (L) for ellipse like precipitate (circle) and twisted diamond shape of precipitate (diamond).}
 \label{elastic_energy_tetragonal}
\end{figure}

\begin{figure}[htbp!]
 \centering
 \includegraphics [width=0.4\linewidth]{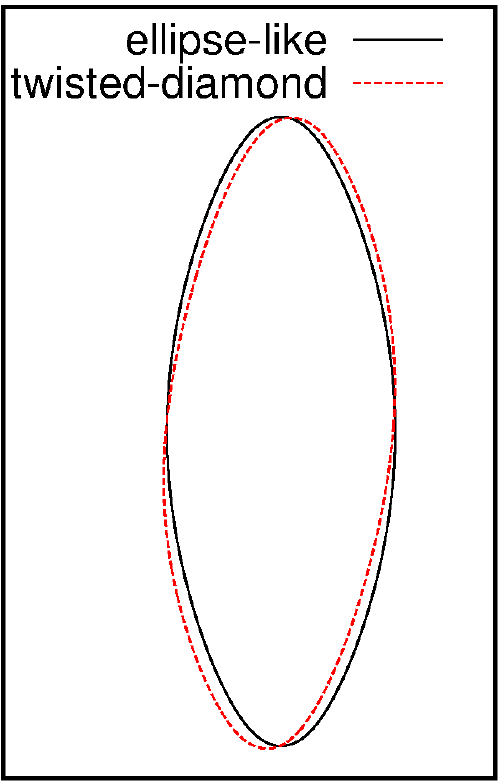} \\
 \vspace{0in}
 \caption{The equilibrium shapes of precipitate with same equivalent 
          radius R=52 (L=3.5), ellipse like shape (thick-dark line) in metastable equilibrium 
          and diamond like shape (dotted red-line) in stable equilibrium.}
 \label{eq_shapes_elastic_energy_check}
\end{figure}

In order to ensure that the phase-field calculations are indeed yielding the lowest
energy shapes, we have computed the equilibrium shapes beyond the bifurcation point which 
are ellipse-like (elongation along y-direction). This is done by starting with an 
initial configuration that is in perfect alignment with the y-axis.
Thereafter, we have calculated corresponding total energies of the precipitates. 
as highlighted in Fig.~\ref{elastic_energy_tetragonal}, 
as function of characteristic length. It is observed that the 
total energy of the twisted diamond like shapes is lower 
than that of the precipitates which are ellipse-like. This indicates that, the precipitates with 
lower energies (twisted diamond shape) are the stabler equilibrium configurations, 
compared to their ellipse-like counterparts, beyond the bifurcation point. Thus as shown in 
Fig.~\ref{eq_shapes_elastic_energy_check}, if we give slight rotation to an ellipse-like 
precipitate (thick dark line), it will acquire a twisted diamond like shape (dotted line) 
as an equilibrium state, beyond the critical point. With increase in the characteristic length 
past the bifurcation point, the difference between the total energies of 
stable and metastable precipitate shapes keeps on increasing. 

\begin{figure}[htbp!]
 \centering
 \includegraphics [width=0.8\linewidth]{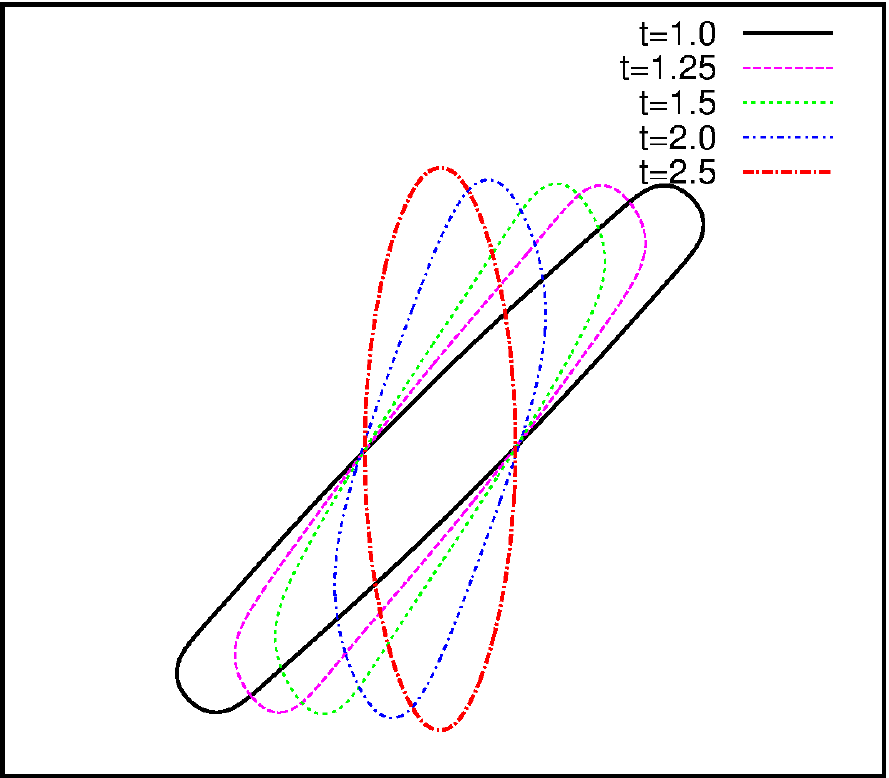} \\
 \vspace{0in}
 \caption{Equilibrium morphologies of precipitate with varying 
          degree of strength of tetragonal misfit which varies from t=1.0 to t=2.5 for Az=0.3 and $\delta=1.0$.}
  \label{tetragonal_eq_shape_tratio}
\end{figure}

As illustrated in the previous section, the change exhibited in the precipitate morphology is an 
effect of the precipitate size alone, where the misfit ratio remains constant i.e. $t=2.0$. Further, 
we also characterize the effect of change in the magnitude of misfit ratio i.e. 
$\epsilon^*_{xx}/\epsilon^*_{yy}$ on an equilibrium morphologies of the precipitate by keeping 
the precipitate size constant. For this purpose, we keep the magnitude of $\epsilon^*_{xx}$ 
constant and varied $\epsilon^*_{yy}$ from 0.004 to 1.0. Accordingly, 
the misfit ratio takes values from 2.5 to 1.0. Fig.~\ref{tetragonal_eq_shape_tratio} shows the 
equilibrium shapes of the precipitate for the different values of misfit ratio with $A_z=0.3$, $\delta=1.0$. 
Here, we retain the same precipitate size for all calculations (i.e. R=65). As discussed earlier, 
with these conditions the precipitate acquires a bifurcated shape which is twisted diamond like. 
With increase in magnitude of misfit ratio, the precipitate orientation changes towards 
the direction of least misfit i.e. the vertical axis. 
Initially, for t=1.0, the equilibrium precipitate is aligned exactly along the elastically softer directions 
which are $<110>$ with an orientation of $45^0$ alongside a rectangular shape 
i.e. in this case the equilibrium shape is determined by the elastically softer directions ($A_z=0.3$) alone. 
As, the misfit ratio elevates from 1.0 to higher values, the precipitate acquires an orientation with 
the larger magnitude, which is an obvious situation, as the orientation is a compromise between the 
tendencies to align along the elastically softer direction ($<110>$) and the 
direction of lower misfit $(010)$. For t=2.5, an equilibrium 
morphology tends towards elongation along the lower misfit directions, 
where the precipitate has an elongated ellipse-like shape aligned along the vertical axis. 
So, the degree of tetragonality increases with increase in the magnitude of misfit ratio, 
i.e. when the difference between the magnitude of $\epsilon^*_{xx}$ and $\epsilon^*_{yy}$ is larger. 
The other way of looking at this is that with 
increase in the value of tetragonality the bifurcation point for the
system shifts to the larger values of equivalent precipitate sizes, 
as the driving force for the precipitate elongating along the direction of lower misfit increases.

\begin{figure}[htbp!]
 \centering
 \includegraphics [width=0.8\linewidth]{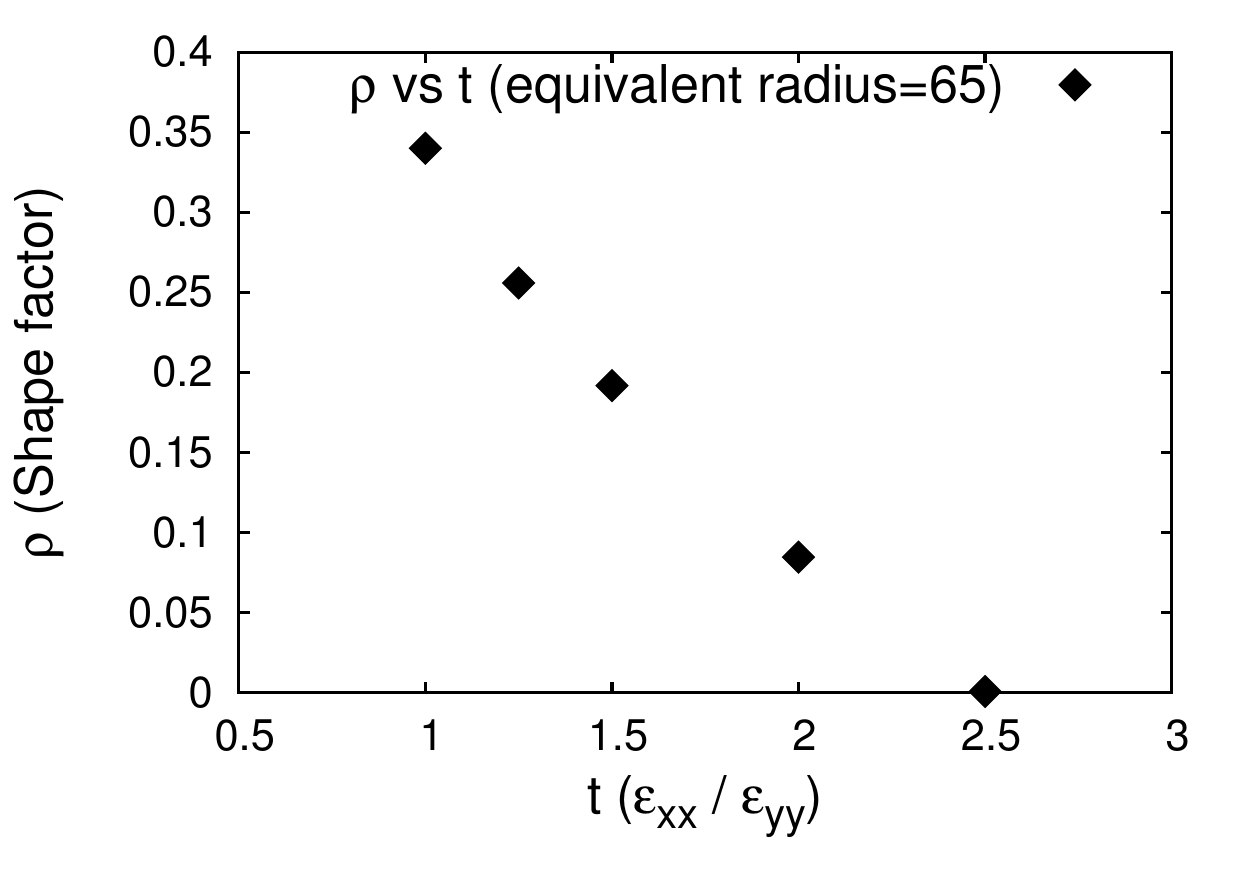} \\
 \vspace{0in}
 \caption{Plot depicts the change in shape factor as function of 
          misfit ratio varying from t=1.0 to t=2.5 for Az=0.3 and $\delta$=1.0.}
  \label{t_vs_rho_tetra}
\end{figure}

\subsubsection{Misfit components with opposite sign}

\begin{figure}[htbp!]
 \centering
 \includegraphics [width=0.35\linewidth]{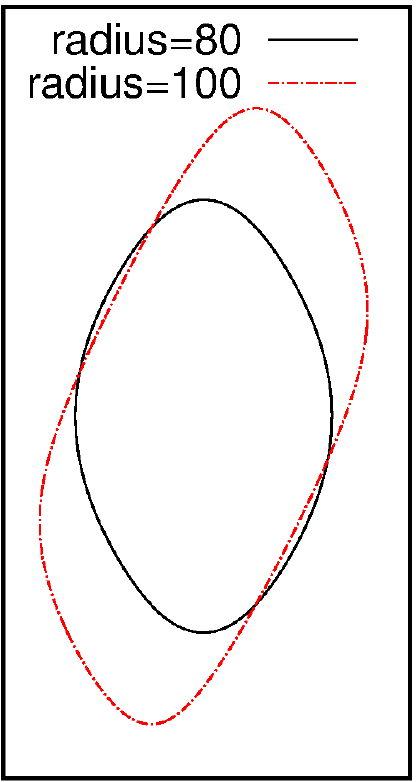} \\
 \vspace{0in}
 \caption{Equilibrium shapes of precipitate with tetragonal misfit and different sizes, 
	  where Az=2.0, t=-2.0, i.e. $\epsilon_{xx}^*=0.01, \epsilon_{yy}^*=-0.005$.}
  \label{tetragonal_eq_shape3} 
\end{figure}

In this case, the misfit components along x and y-directions have opposite sign as well as different 
magnitude. Here, we have chosen $t=-2.0$ i.e. $\epsilon_{xx}^*=0.01, \epsilon_{yy}^*=-0.005$ and $A_z=2.0$. 
Although, the symmetry of misfit is the same as compared with the previous section, there is 
an important difference. As shown in Fig.~\ref{tetragonal_eq_shape3}, there is a shape transition in this case too,
where a precipitate with size R=80, has an ellipse like shape, while above the critical characteristic 
length, precipitate (R=100), a twisted diamond like shape results as the equilibrium 
morphology. This shape transition is seen in all the cases where 
$A_z<1.0, A_z=1.0$ and $A_z>1.0$. So, the shape bifurcation occurs even at $A_z=1.0$ and 
$A_z>1.0$, which is in contrast to the previous case where the principal components of misfit 
strains have the same sign. 

\begin{figure}[htbp!]
 \centering
 \includegraphics [width=0.35\linewidth]{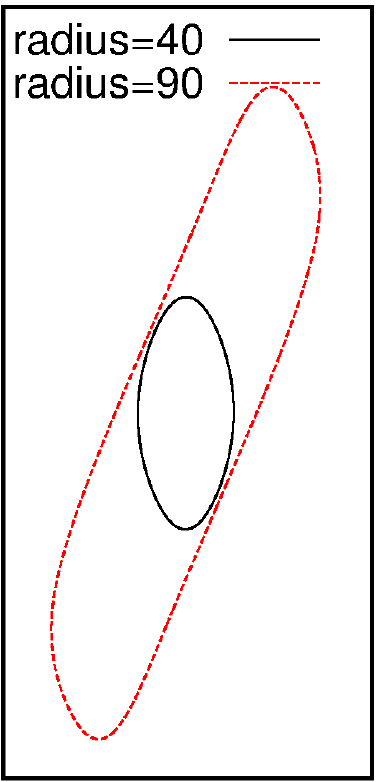} \\
 \vspace{0in}
 \caption{Equilibrium shapes of precipitate with tetragonal misfit and different sizes, 
	  where Az=0.3, t=-2.0, i.e. $\epsilon_{xx}^*=0.01, \epsilon_{yy}^*=-0.005$.}
  \label{tetragonal_eq_shape4} 
\end{figure}

Fig.~\ref{tetragonal_eq_shape4}, also shows that the precipitate with smaller radius (R=40) 
and $A_z$ is less than one ($A_z=0.3$), 
has an equilibrium shape which has its axis elongated along the direction of lower misfit. 
For the larger precipitate shapes (R=90), we acquire a twisted 
diamond like shape. It is observed that there is an influence of the change in 
magnitude of inhomogeneity moduli 
($\delta$) and the magnitude of Zener anisotropy parameter ($A_z$) on the critical characteristic 
length ($L_c$) of shape bifurcation. For a systematic analysis we keep $\delta$ constant 
($\delta=1.0$) i.e. the homogeneous moduli, while the Zener anisotropy parameter ($A_z$) is varied 
from 0.3 to 2.0. We noticed that, the $L_c$ is lower for $A_z=0.3$ compared to the case of $A_z=2.0$. 
For $A_z=0.3$, the precipitates with smaller sizes acquire a twisted diamond like shape, 
as $<110>$ are elastically softer directions. This provides a driving force for the precipitates 
even with the smaller sizes to orient along the elastically softer directions. 
In contrast for $A_z>1.0$ (in this case $A_z=2.0$), the elastically soft direction are $<100>$ 
which is also the same as the direction of lowest misfit. Thus, it becomes 
harder for the precipitates to orient along $<110>$ direction and acquire the twisted diamond 
like shape. This is evident from Fig.~\ref{tetragonal_eq_shape3}, where we also find 
that the critical characteristic length ($L_c$) is pushed to larger 
values i.e. the shape transition occurs for the larger precipitate sizes. 

\subsection{Competition between anisotropy in interfacial energy and elastic energy}

This section illustrates two factors controlling the selection of orientation of 
the precipitate morphology i.e. the anisotropies in interfacial energy ($\varepsilon$) and 
elastic energy ($A_z$). In order to investigate the influence of this competitive 
nature of the energy anisotropies, we first vary the anisotropy in elastic energy 
by changing the magnitude of $A_z$ (with constant tetragonality, i.e. t=+1.0), while 
keeping the anisotropy in interfacial energy constant and then repeat the other way around. 
This is achieved by incorporating anisotropy in interfacial energy as elaborated in ~\cite{karma1998}: 
\begin{align}
 \gamma &= \gamma_0 a(\bm{n}), \\ \nonumber
 \gamma &= \gamma_0 \left(1 - \varepsilon\left(3 -4\dfrac{{\phi}^{'4}_{x} + {\phi}^{'4}_{y}}{({\phi}^{'2}_{x} + {\phi}^{'2}_{y})^2} \right) \right),
\end{align}
where, $\varepsilon$ is the strength of anisotropy in interfacial energy, $a(\bm{n})$ is the anisotropy function
of the interface normal $\bm{n}$, which has components $\bm{n_x}=-\dfrac{{\phi}^{'}_{x}}{|\nabla \phi|}$, 
$\bm{n_y}=-\dfrac{{\phi}^{'}_{y}}{|\nabla \phi|}$, with ${\phi}^{'}_{x}$
and ${\phi}^{'}_{y}$ being the partial derivatives of the order parameter $\phi\left(x,y\right)$
in the x and y-directions. 

\subsubsection{Effect of anisotropy in elastic energy ($A_z$)}
\begin{figure}[htbp!]
 \centering
 \includegraphics [width=0.5\linewidth]{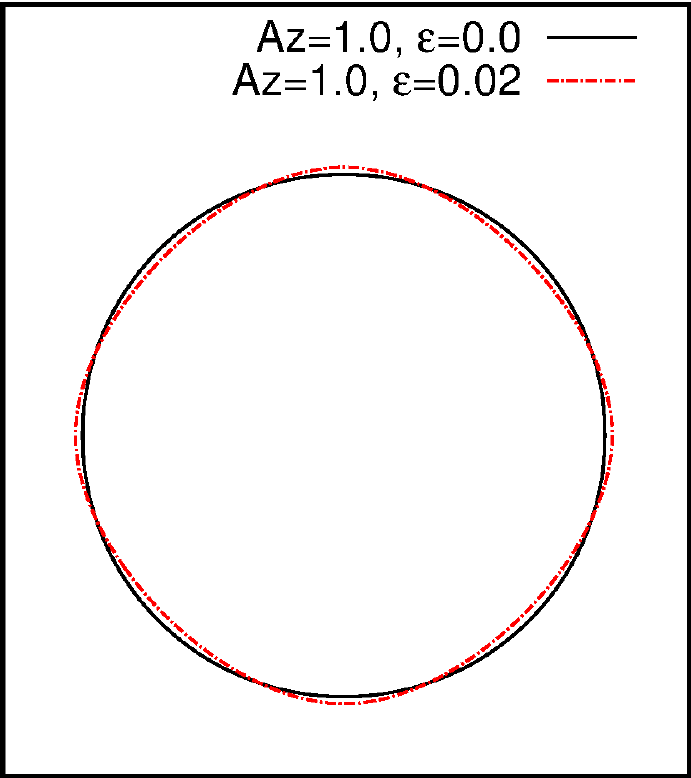} \\
 \vspace{0in}
 \caption{Equilibrium shapes of precipitate with (red dotted line) 
  and without (thick dark line) anisotropy in interfacial energy, for Az=1.0, t=1.}
 \label{int_aniso_az1}
\end{figure}

\begin{figure}[htbp!]
 \centering
 \includegraphics [width=0.5\linewidth]{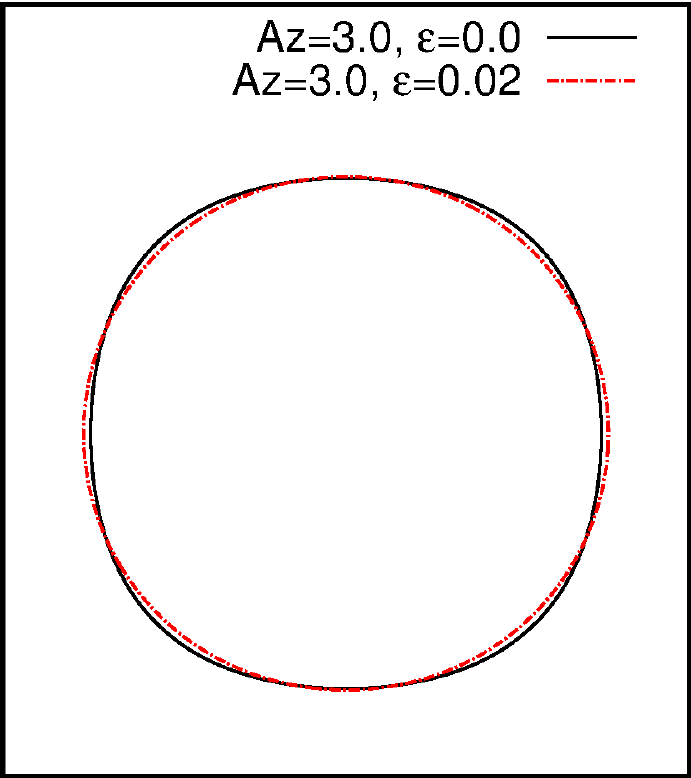} \\
 \vspace{0in}
 \caption{Equilibrium shapes of precipitate with (red dotted line) 
  and without (thick dark line) anisotropy in interfacial energy, for Az=3.0, t=1.}
 \label{int_aniso_az3}
\end{figure}

\begin{figure}[htbp!]
 \centering
 \includegraphics [width=0.5\linewidth]{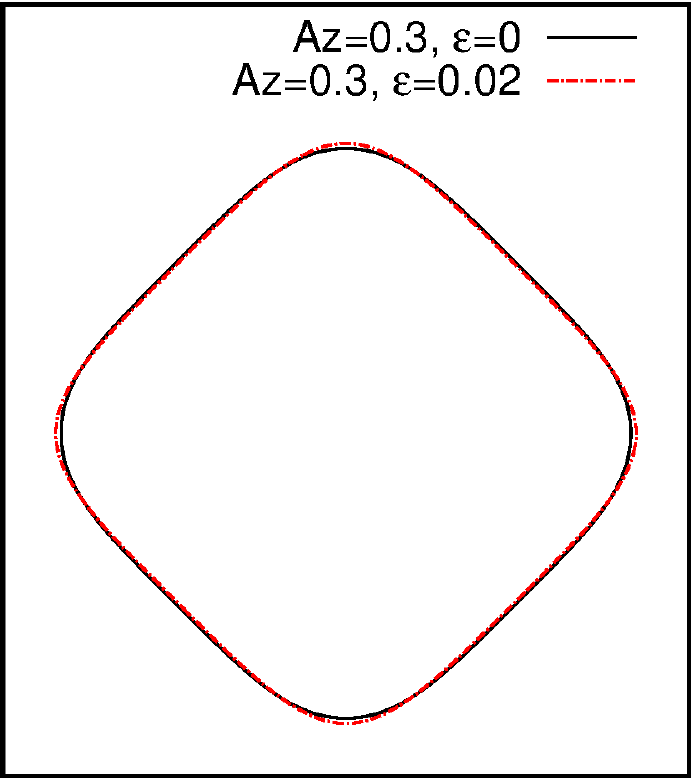} \\
 \vspace{0in}
 \caption{Equilibrium shapes of precipitate with (red dotted line) 
	  and without (thick dark line) anisotropy in interfacial energy, for Az=0.3, t=1.}
 \label{int_aniso_az0.3}
\end{figure}

Here, the magnitude of $A_z$ is altered from 0.3 to 3.0 for a constant $\varepsilon=0.02$. 
Fig.~\ref{int_aniso_az1} -~\ref{int_aniso_az0.3}, show the equilibrium precipitate morphologies 
with (dotted line) and without (thick line) anisotropy in interfacial energy, with increasing the 
strength of $A_z$. In all the cases, the size of precipitate is kept same i.e. the equivalent radius is 
constant ($R=25$). In this regard, three prominent cases are investigated i.e. $A_z=0.3, Az=1.0$ and $A_z=3.0$, 
where the influence of the change in magnitude of $A_z$ on an equilibrium morphology is discussed. 
All these simulations are performed with the precipitate sizes which are well below the bifurcation point. 

In the first case, we keep $A_z=1.0$, with $\varepsilon=0.02$. 
This is to quantify the effect of anisotropy in interfacial energy alone. This is summarized in 
Fig.~\ref{int_aniso_az1}, where the precipitate takes a circular shape (thick dark line) due to 
isotropic elastic energy with no anisotropy in interfacial energy. With the introduction of 
anisotropy in interfacial energy, the precipitate turns to a diamond like shape, with its faces parallel to $<110>$ 
directions. So, the equilibrium shape of the precipitate is primarily 
determined by interfacial anisotropy alone. 

Fig.~\ref{int_aniso_az3}, shows an equilibrium precipitate morphology (thick dark line) 
with $A_z=3.0$, where there is a cubic anisotropy in the elastic energy which drives the precipitate 
to acquire a square like shape with rounded corners where the square faces are aligned 
along the elastically soft directions $<100>$. Again, incorporating anisotropy in 
interfacial energy ($\varepsilon=0.02$), 
tends to align the precipitate faces along $<110>$ directions. 
This minimizes the effect of elastic anisotropy on precipitate shape, by giving rise 
to a morphology which takes a shape in between a square and a diamond. 
This is shown in Fig.~\ref{int_aniso_az3}, 
where the equilibrium shape of the precipitate (dotted line) has acquired a shape as explained above. 
There can be a possibility, where effect of both anisotropies counterbalance 
each other and gives rise to a morphology that is circular, 
though the respective energies (elastic and interfacial) possess anisotropies.

When $A_z<1$, (in this case $A_z=0.3$), the precipitate acquires a shape which prefers 
to align its face along $<110>$ directions that is elastically softer. This is captured 
in Fig.~\ref{int_aniso_az0.3}, which shows the precipitate morphologies where both 
the interfacial and elastic anisotropies drive the precipitate shape towards a diamond.
Precipitate without anisotropy in interfacial energy, has its faces parallel to $<110>$ directions, 
whereas with anisotropy in interfacial energy the precipitate faces becomes weakly concave towards the center of 
the precipitate with the corners becoming more sharp. It is evident that the presence of interfacial 
anisotropy along with cubic anisotropy in elastic energy affects equilibrium morphologies 
of the precipitate distinctly. 

\begin{figure}[htbp!]
 \centering
 \includegraphics [width=0.5\linewidth]{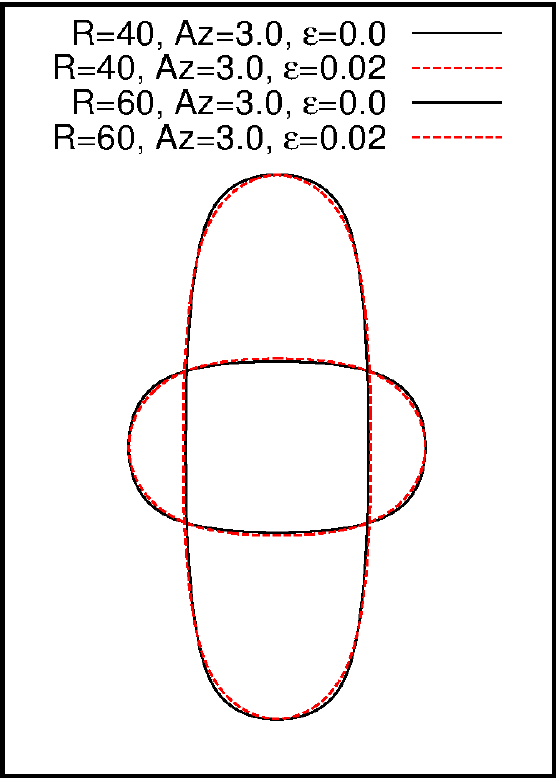} \\
 \vspace{0in}
 \caption{Equilibrium shapes of precipitate beyond bifurcation point 
	  with (red dotted line) and without (thick dark line) anisotropy in interfacial energy, for Az=3.0, t=1.}
 \label{eq_shapes_az3_sf2_bfr}
\end{figure}

In the preceding discussion, we characterize the effect of energy anisotropies on the precipitate 
shape where the precipitate size is well below the bifurcation point ($L_c$). 
Now, we shift beyond the bifurcation point ($L_c$) with the same argument i.e. competitive 
effect of both the anisotropies while accessing the larger precipitate sizes. 
Fig.~\ref{eq_shapes_az3_sf2_bfr} shows such morphologies of the precipitate 
with $A_z=3.0$. Precipitates with two different sizes (R=40 and R=60) and different 
variants are shown. Presence of anisotropy in interfacial energy does not affect the equilibrium morphology 
of precipitate significantly, whereby, though the precipitate corners become more rounded, edges of 
the precipitate remain strongly aligned along the elastically softer directions ($<100>$). This implies 
that with increase in the size of precipitate, the influence of elastic energy anisotropy 
dominates over interfacial energy anisotropy.

\subsubsection{Effect of anisotropy in interfacial energy ($\varepsilon$)}
\begin{figure}[!htbp]
 \centering
 \subfigure[]{\includegraphics[width=0.5\linewidth]{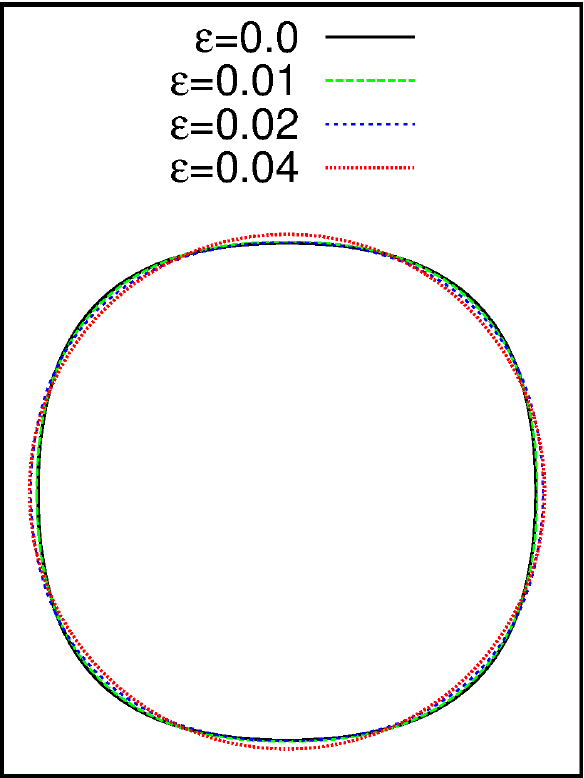}
 \label{az3_diff_sfaniso}
 }
\caption{The variation of equilibrium shapes of precipitate as a function of strength of interfacial anisotropy 
	 where $\varepsilon$ varies from 0 to 0.04 and Az=3.0, t=1.}
\label{eq_shapes_az3_vs_sf_anisotropy}
\end{figure}

While in the previous section, the strength of anisotropy in interfacial energy ($\varepsilon$) 
is kept constant, furthering the discussion, in this section, 
we will consider the effect of varying $\varepsilon$ on the precipitate morphology. Here, we vary 
the magnitude of $\varepsilon$ from 0.0 to 0.04, while holding the Zener anisotropy parameter constant, 
at a value of $A_z=3.0$. The equilibrium morphology of the precipitate initially acquires a cubic shape with 
rounded corners, while the precipitate faces are aligned along the elastically soft directions ($<100>$). 
Further, with increase in the strength of $\varepsilon$, the precipitate faces start to orient towards 
$<110>$ directions. This is due to a significant increase in the strength of anisotropy in interfacial 
energy, while keeping the magnitude of $A_z$ constant. Thus, an increase in the strength of 
anisotropy in interfacial energy imparts a driving force for alignment of the 
precipitate faces along $<110>$ directions rather than $<100>$ directions 
(elastically softer), giving rise to a diamond like shape as shown 
in Fig.~\ref{az3_diff_sfaniso}. 

\begin{figure}[htbp!]
 \centering
 \includegraphics [width=1.0\linewidth]{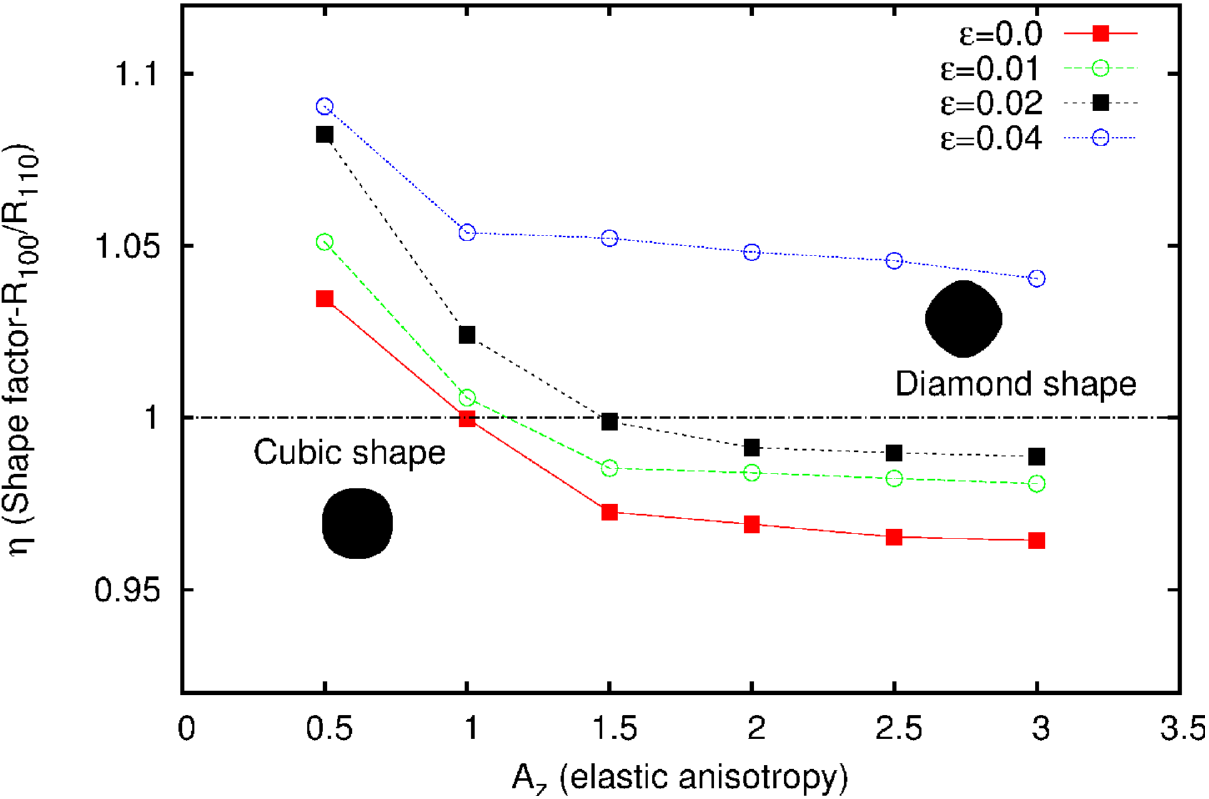} \\
 \vspace{0in}
 \caption{Map represents the shape factor ($\eta$) as function of $A_z$ for different strengths of interfacial anisotropy, 
	  dark horizontal line separates the region between cube(square) like shapes of precipitate 
	  and diamond like shapes of precipitates which are characterized by $\eta$.}
 \label{az_vs_shape_factor_for_diff_sfaniso}
\end{figure}

The characterization of change in the precipitate morphology is precisely captured by plotting 
the shape factor ($\eta$) as a function of the strength of elastic anisotropy for a given strength of 
anisotropy in interfacial energy, which is shown in Fig.~\ref{az_vs_shape_factor_for_diff_sfaniso}. 
Here, the shape factor is defined as the ratio 
of the precipitate size along $<100>$ to size along $<110>$. Significance of calculating 
$\eta$ in this way is that, it determines the precipitate orientation. If $\eta < 1.0$, 
it implies that the precipitate faces are aligned along the elastically soft directions i.e. $<100>$ directions 
and the anisotropy in elastic energy is the shape determining factor provided $A_z>1.0$. 
Similarly, if $\eta > 1.0$, it implies that the precipitate faces preferentially align along 
$<110>$ directions which is determined by 
the anisotropy in interfacial energy and $\eta=1.0$ gives circular shape of the precipitate, 
where there are no concavities in the morphology. 
Fig.~\ref{az_vs_shape_factor_for_diff_sfaniso}
depicts a morphology map, which plots the shape factor as a function of the 
different anisotropy strengths in the elastic energy $A_z$, while considering
different values of anisotropy in interfacial energy $\varepsilon$. 
While for the case where $\varepsilon\leq0.02$,
there is a critical value beyond which the shape transition occurs from 
a diamond to a cube, for the larger values of anisotropy in interfacial energy, the equilibrium 
shape remains diamond like. 
The morphology map depicts that the morphology of the precipitate corresponding 
to $A_z=1.5$ and $\varepsilon=0.02$ acquires a near circular shape. 
However, this transition point is a function of the size of the precipitate. 

\section{Comparison with previous models}
In this section, we compare our diffuse-interface model against 
the previous/available models for the calculation of 
equilibrium morphologies of precipitates. Here, as a starting 
point it is useful to analyze the sharp-interface limit of
the model as this will enable comparison with other models which 
are either based on the level-set algorithm, FEM,boundary-integral 
methods or diffuse-interface methods.

For this, we recall the evolution equation, as in Eqn.\ref{phi_evolve}, 
written for simplicity for the case of isotropic interfacial energy.
We transform this equation into a co-ordinate system that is normal 
to the interface at a given location where the order parameter is being updated.
Thus, the interface normal $\nabla \phi$ can be written as,
\begin{align}
 \nabla\phi = \mathbf{n}\dfrac{\partial \phi}{\partial n},
\end{align}
which upon taking a divergence yields,

\begin{align}
 \nabla^{2}\phi &= \dfrac{\partial^{2}\phi}{\partial n^{2}} + \dfrac{\partial \phi}{\partial n} \left(\nabla\cdot \mathbf{n}\right).
\end{align}

Therefore, the transformed equation reads, 

\begin{align}
 \tau W \dfrac{\partial \phi}{\partial t} &= 2\gamma W \left(\dfrac{\partial^{2}\phi}{\partial n^{2}} + \dfrac{\partial \phi}{\partial n} \left(\nabla\cdot \mathbf{n}\right)\right)
                                           - \dfrac{16}{\pi^{2}}\dfrac{\gamma}{W}\left(1-2\phi\right) - \nonumber\\ 
                                           &\dfrac{\partial f_{el}}{\partial \phi} - \lambda_{\beta}h'\left(\phi\right).
\end{align} 

The leading order solution for the phase-field profile $\phi^{0}$ satisfies the equation, 

\begin{align}
 2\gamma W \dfrac{\partial^{2}\phi^{0}}{\partial n^{2}} - \dfrac{16}{\pi^{2}}\dfrac{\gamma}{W}\left(1-2\phi^{0}\right) = 0,
 \label{equi_partition}
\end{align}

which upon integration once along the interface normal along 
with the boundary conditions that in the bulk precipitate and the 
matrix $\dfrac{\partial \phi^{0}}{\partial n}=0$, gives, 

\begin{align}
 \gamma W \left(\dfrac{\partial \phi^{0}}{\partial n}\right)^{2} =  \dfrac{16}{\pi^{2}}\dfrac{\gamma}{W}\phi^{0}\left(1-\phi^{0}\right), 
\end{align}

the solution to this is also the equilibrium phase-field profile without 
any driving forces. Along with this the leading order solutions to the 
stress-strain profiles are determined from mechanical equilibrium
conditions, in the limit that the interface thickness tends to zero.
Without extensive math and drawing from the results of the phase-field
simulations which deliver the stress-strain profiles along the 
interface normal(Appendix ~\ref{1d_stress-strain}), we see that in the limit of vanishing interface
thickness, where the values of the properties at the interface and the asymptotic extensions
from the bulk on either side onto the interface become the same, 
the following conditions will be satisfied, 

\begin{align}
 \sigma_{nn}^{\alpha} = \sigma_{nn}^{\beta} = \sigma_{nn}^{0},\\ \nonumber
 \sigma_{nt}^{\alpha} = \sigma_{nt}^{\beta} = \sigma_{nt}^{0},
\end{align}

which is also the consistent with the 
condition for zero traction along the normal, while the strains,

\begin{align}
 \epsilon_{tt}^{\alpha} = \epsilon_{tt}^{\beta} = \epsilon_{tt}^{0},
\end{align}

are continuous across the interface. The other components
however, are going to exhibit a jump across the interface.
These are same jump conditions that we utilize in the tensorial 
interpolation scheme and also elaborated in ~\cite{schneider2015}.
Using the preceding arguments the 
leading order phase-field update derives as, 

\begin{align}
 \tau W \dfrac{\partial \phi}{\partial t} &= 2\gamma W \dfrac{\partial \phi^{0}}{\partial n} \left(\nabla\cdot \mathbf{n}\right)
                                           - \dfrac{\partial f_{el}}{\partial \phi} - \lambda_{\beta}h'\left(\phi\right).
\label{phi_leading_order}
\end{align}

Additionally, using the sharp-interface limits of the stress and strain profiles, 
we can also derive the following, 

\begin{align}
 \dfrac{\partial f_{el}}{\partial \phi} &= \dfrac{df_{el}}{d\phi} - \dfrac{\partial f_{el}}{\partial \epsilon_{nn}}\dfrac{\partial \epsilon_{nn}}{\partial \phi}
                                        -2\dfrac{\partial f_{el}}{\partial \epsilon_{nt}}\dfrac{\partial \epsilon_{nt}}{\partial \phi}
                                        -\dfrac{\partial f_{el}}{\partial \epsilon_{tt}}\dfrac{\partial \epsilon_{tt}}{\partial \phi}.
\end{align}

Since, in the sharp-interface limit, the strain-component $\epsilon_{tt}$ is continuous
across the interface the last differential in the preceding equation vanishes, 
in the limit that the interface thickness becomes zero, thereby, the leading 
order elastic driving force derives as, 

\begin{align}
 \dfrac{\partial f_{el}}{\partial \phi} &= \dfrac{df_{el}}{d\phi} - \sigma_{nn}^{0}\dfrac{\partial \epsilon_{nn}}{\partial \phi} -2\sigma_{nt}^{0}\dfrac{\partial \epsilon_{nt}}{\partial \phi}\nonumber\\
                                        &= \dfrac{d}{d\phi}\left(f_{el} -\sigma_{nn}^{0}\epsilon_{nn}-2\sigma_{nt}^{0}\epsilon_{nt}\right).
\end{align}

Substituting in the leading-order phase-field update as in Eqn.\ref{phi_leading_order}, 
we derive,

\begin{align}
 \tau W \dfrac{\partial \phi}{\partial t} &= 2\gamma W \dfrac{\partial \phi}{\partial n} \left(\nabla\cdot \mathbf{n}\right)
                                           - \dfrac{d}{d\phi}\left(f_{el} -\sigma_{nn}^{0}\epsilon_{nn}-2\sigma_{nt}^{0}\epsilon_{nt}\right) \nonumber\\
                                           &- \lambda_{\beta}h'\left(\phi\right).
\end{align}

Multiplying both sides of the preceding equation with the leading order phase-field profile, 
$\dfrac{\partial \phi^{0}}{\partial n}$, and integrating from inside the bulk precipitate
into the matrix in the normal direction, we derive, 

\begin{align}
 \int_{in}^{out}\tau W \dfrac{\partial \phi}{\partial t}\dfrac{\partial \phi^{0}}{\partial n} dn 
                                                              &=  2\gamma W \int_{in}^{out}\left(\dfrac{\partial \phi^{0}}{\partial n}\right)^{2} \left(\nabla\cdot \mathbf{n}\right)dn\nonumber\\
                                                             -& \int_{1}^{0}\dfrac{d}{d\phi}\left(f_{el} -\sigma_{nn}^{0}\epsilon_{nn}-2\sigma_{nt}^{0}\epsilon_{nt}\right)d\phi\nonumber\\
                                                              &- \int_{1}^{0}\lambda_{\beta}h'\left(\phi\right)d\phi.
\end{align}

Using the property that, 
$\dfrac{\partial \phi}{\partial t}\dfrac{\partial \phi^{0}}{\partial n} = - \left(\dfrac{\partial n}{\partial t}\right)_{\phi} \left(\dfrac{\partial \phi^{0}}{\partial n}\right)^{2}=
-v_n\left(\dfrac{\partial \phi^{0}}{\partial n}\right)^{2}$, where additionally $\left(\dfrac{\partial n}{\partial t}\right)_{\phi}$
is replaced with the normal velocity $v_n$. Further, $\nabla\cdot \mathbf{n} = -\kappa$ (ignoring changes of curvature along the interface normal) 
thereby, the following equation is derived, 

\begin{align}
 -\tau W v_n \int_{in}^{out}  \left(\dfrac{\partial \phi^{0}}{\partial n}\right)^{2} dn  &= - 2\gamma W \kappa \int_{in}^{out} \left(\dfrac{\partial \phi^{0}}{\partial n}\right)^{2}dn\nonumber\\
                                                                                       -&  \left(f_{el} - \sigma_{nn}^{0}\epsilon_{nn} -2\sigma_{nt}^{0}\epsilon_{nt}\right)|_{1}^{0}\nonumber\\
                                                                                       &- \lambda_{\beta}h\left(\phi^{0}\right)|_{1}^{0}.
\end{align} 

Using, the relation, in Eqn.\ref{equi_partition}, we can integrate
 $\int_{in}^{out} \left(\dfrac{\partial \phi^{0}}{\partial n}\right)^{2} dn$
as $\int_{1}^{0}  \left(\dfrac{\partial \phi^{0}}{\partial n}\right) d\phi$, or,
-$\int_{1}^{0}  \dfrac{16}{\pi^{2}}\sqrt{\phi^{0}\left(1-\phi^{0}\right)} d\phi^{0}$, 
that returns $\dfrac{1}{2W}$. Combining, the results into the preceding equation, 
we derive, 

\begin{align}
 \dfrac{1}{2}\tau v_n &= \gamma \kappa  + \left(\omega_\beta - \omega_\alpha\right) - \lambda_{\beta},
\end{align}

where we have assigned, $\omega = \left(f_{el} - \sigma_{nn}^{0}\epsilon_{nn} -2\sigma_{nt}^{0}\epsilon_{nt}\right)$.
The preceding sharp-interface limit is a relation between the 
velocity of advance of the interface front as result of the capillarity 
forces $\gamma \kappa$ as well as the elastic configurational forces
$ \left(\omega_\beta - \omega_\alpha\right)$, while the Lagrange-parameter
$\lambda_{\beta}$ maintains the volume constraint. This is similar to the
condition that is derived by Zhao et al.\cite{zhao_duddu_bardas2013, zhao_bardas2015},
wherein, $ \left(\omega_\beta - \omega_\alpha\right)$ is nothing but the 
term $\mathbf{n}\cdot\left(\Sigma_\beta - \Sigma_\alpha\right)\cdot \mathbf{n}$, 
with $\Sigma$ being the Eshelby momentum tensor \cite{eshelby1975} 
$f_{el} I  - \sigma\cdot\nabla \mathbf{u}$ ($\mathbf{u}$ is the displacement 
field), that is used in the paper.
Similarly, for the condition of equilibrium, $v_n$ is identically
zero at all the interface points for which we have, 

\begin{align}
  \lambda_{\beta} = \gamma \kappa  + \left(\omega_\beta - \omega_\alpha\right).
\end{align}

Comparing this with the balance condition that is utilized in the work 
by Thomson and Voorhees \cite{voorhees1992}, to generate the set of 
integro-differential equations for the solutions of the interface 
co-ordinates, $\lambda_\beta$ is nothing but the departure of the 
chemical potential between the phases $\alpha$ and $\beta$ at a
constant diffusion potential. This is also remarked in Schmidt
and Gross \cite{schmidt1997}, although the algorithms of 
arriving at the interface solutions are different. Also, for
the more generic case in the presence of interfacial
energy anisotropy, $\gamma\kappa$ should be replaced with 
$\nabla\cdot\xi$, where $\xi$ is the Cahn-Hoffmann 
zeta vector \cite{cahn_hoffmann1974} as in \cite{voorhees1992}, 
and which can be also shown in the present phase-field 
model in the sharp-interface limit. The advantage in our 
phase-field model is that with just a slightly additional 
cost of resolving the diffuse-interface model, complicated
shapes may be simulated in all dimensions 
with quite simple discretization 
techniques than the ones employed in the FEM as well 
as level-set methods. A more subtle point but noteworthy, 
is that the term $\left(\omega_\beta - \omega_\alpha\right)$
is exactly the same that results from $\dfrac{\partial f_{el}}{\partial \phi}$ 
in the tensorial interpolation 
scheme that we describe in the section on the 
model formulation. The tensorial interpolation exactly 
utilizes the jump conditions of the stress and the strain 
profiles in the sharp-interface limit in the interpolation of
the elastic energies of the individual phases. 

Our present formulation of the phase-field model is in 
contrast to other contemporary models such as 
in \cite{lowengrub_leo1998, wang_Khachaturyan1995, wang_khachaturyan1998,
voorhees_jokisaari2017}, that use the modified Cahn-Hilliard
equation, incorporating for elasticity. In these models, 
the misfit strains are enslaved as functions of the composition 
field, along with the values of the stiffness tensor across
the interface. The equilibrium morphology is derived by solving the 
modified Cahn-Hilliard evolution equation until the
effective chemical potential is the same everywhere.
In terms of computational efficiency this is more costly 
than our model because our evolution equation concerns 
the update of the order parameter $\phi$ whose variation, 
is only over a finite length defined by the diffuse-interface
region, that is much smaller in extent compared to the 
corresponding solution of the composition field over the 
entire integration domain. Apart from this, there is also
a physical difference which stems from the fact that our 
misfits and stiffness tensors are enslaved to the value of the order parameter
$\phi$. This implies that the condition of diffusional 
and mechanical equilibrium may be decoupled
as is the situation in the classical Johnson and Cahn 
model \cite{johnson1984elastically} or the numerical 
FEM models \cite{voorhees1992, schmidt1997, lowengrub_leo1998,jog2000symmetry,
zhao_duddu_bardas2013}, where the composition in the matrix 
and precipitate satisfy just the Laplace equation with the 
constraint of the same diffusion potential. However, 
for the diffuse-interface models that use the composition to parameterize the 
variation of the misfits and stiffness across the interface, where the 
coupling of the diffusional equilibrium and the compositional equilibrium, 
entails that, just the chemical potential $\mu$ arising from the 
chemical part of the free-energy is not constant in the domain, whereas, 
it is the effective chemical potential $\mu-W\nabla^{2}c - \sigma:\delta\epsilon^{*}h'(c)
+\dfrac{1}{2}\delta \sigma:\left(\epsilon-\epsilon^{*}\left(c\right)\right)g'(c)$
that is same everywhere, where $\delta \epsilon^{*}$ is the difference
of the eigenstrain between the phases, while  $h(c)$ and $g(c)$
are interpolation functions of the scaled compositions, which smoothly 
go to fixed values in the bulk phases. $\delta \sigma=
\left(C^{\alpha}-C^{\beta}\right)\left(\epsilon-\epsilon^{*}\left(c\right)\right)$, 
where $\epsilon^{*}\left(c\right)$ is the interpolated eigenstrain.
The interpolation functions $h\left(c\right)$
and $g\left(c\right)$ are typically chosen such that 
in the bulk their derivatives return zero, that essentially
implies that the modification to the boundary condition of the 
diffusion potential w.r.t the sharp-interface problem is only 
at the interface. In the classical Cahn-Hilliard problem without
elasticity this variation of the diffusion-potential leads
to a grand-potential excess at the interface, which 
when integrated across the interface determines the interfacial 
energy $\gamma$ in the system. In the presence of elastic
energies the variation of the diffusion potential in the 
interface is modified and thereby also the interfacial 
energy. The interfacial energy therefore scales
with the interface width $W$ while being a function of the 
stress and strain distribution 
in the interface. Additionally, the interfacial 
compositions as well as the effective misfits and stiffness 
will exhibit a change along the interface between the 
precipitate and the matrix, as the stresses vary with the 
direction of the normal to the interface. As argued by Leo 
et al.\cite{lowengrub_leo1998}, such 
modifications scale with the value of the interface width $W$ 
and in the limit of vanishing interface thickness the 
sharp-interface problem is retrieved. It is also important to 
mention that in our model formulation as well, the Khachaturyan
type interpolation does yield an interfacial excess which modifies
the interfacial energy from the model parameter $\gamma$ 
and the deviation scales with $W$, although the nature of the excess
is different. This interfacial 
excess is removed with the tensorial interpolation as described in this 
paper and in ~\cite{schneider2015} as it respects the 
precise jump conditions corresponding to the 
sharp-interface problem. However, given the 
subtle but important conceptual difference between our model 
formulation and the other diffuse-interface models, it is presently 
unclear to us, which amongst the modeling schemes, will yield 
acceptable errors with respect to the sharp-interface problem, 
with the larger interface width $W$.

With regards to the actual situation experimentally, 
the misfit may truly be a strong function of the composition.
The problem is even quite complicated in 
multi-component alloys, where given the different possible equilibria, 
the interesting situation is that the choice of the equilibrium compositions, 
controls the magnitude of the misfit and thereby also the shape. 
The three-factors are coupled and thereby it is unlikely that 
the misfit parameterization can be done as a function of the 
order-parameter as performed in our model presentation or held as a
constant value as in the classical analytical and FEM models. Consequently, 
the mechanical equilibrium and the equilibration of the composition
field may need to be performed in a coupled fashion along with the 
solution to the Allen-Cahn equation for the order-parameter update under 
the constraint of constant volume. This is similar to the 
chemo-mechanical model level-set based FEM approach 
proposed by Zhao \cite{zhao_bardas2015}, 
which also incoporates interfacial stresses. Extension of our 
model along these lines particularly for addressing the situations
of multi-component alloys is possible. 

\section{Conclusion}

In this paper, we have presented a phase-field model 
for the determination of equilibrium morphology of precipitates
under coherency stresses. Our model
couples the Allen-Cahn dynamics of the order-parameter
evolution with the constraints of mechanical equilibrium as 
well as volume constraint. Using this model, we predict the 
symmetry breaking transitions and bifurcation diagrams which occur 
for the shapes of the precipitates as a function of size, for 
various combinations of misfits as well as anisotropies in the 
elastic constants. These predictions are compared against 
available analytical solutions ~\cite{johnson1984elastically} as well as sharp-interface 
computations ~\cite{jog2000symmetry} where we find excellent agreement.
In the system with non-dilatational misfits, when misfits along principal 
directions have same sign, the shape transition (ellipse to twisted diamond shape) 
is observed only when $A_z<1$, otherwise the precipitate keeps elongating along the axis 
of direction with lower misfit. When misfits along principal directions have opposite 
signs, the precipitate shape transition occurs irrespective to the magnitude of $A_z$. 
These results are in very good agreement with previous numerical results
~\cite{jog2000symmetry},~\cite{Sankarasubramanian2002}.
Along, with this we also study the combined influence
of elastic and interfacial energy anisotropy both above
and below bifurcation. Here, we find that in the presence
of interfacial energy anisotropy, there is a continuous change
in the shape of the precipitate with size before bifurcation, 
which is different from the case where there is just elastic 
anisotropy. Apart from this we also compare and contrast 
our model formulation with other diffuse-interface
and sharp-interface models. Here, while we discuss that 
the sharp-interface model descriptions are retrieved in the 
different models in the sharp-interface limit, 
we also highlight the conceptual differences
between our model formulation and the other diffuse-interface
methods. Amongst the future outlooks,
is the coupling of the Allen-Cahn dynamical equation with 
a time-dependent diffusional equation for the study of 
growth and coarsening of precipitates in multi-component 
systems.

\section{Appendix}
\subsection{Elastic free energy density}
\label{prefactors}
Eqn~\ref{fel_phi} gives the elastic free energy density, which includes several prefactors i.e. $Z_3, Z_2, Z_1, Z_0$. 
These prefactors are dependent on particular values of elastic constant in respective phases 
i.e. the precipitate and matrix. Their expressions are as follows:
Here $C^{\alpha,\beta}_{11} = C^{\alpha,\beta}_{1111}, C^{\alpha,\beta}_{22} = C^{\alpha,\beta}_{2222}, 
C^{\alpha,\beta}_{44} = C^{\alpha,\beta}_{1212}, C^{\alpha,\beta}_{12} = C^{\alpha,\beta}_{1122}$.
\begin{align*}
 Z_3 &= \left(C^{\alpha}_{11}-C^{\beta}_{11}+C^{\alpha}_{12}-C^{\beta}_{12}\right)\epsilon^{*2},\\
 Z_2 &= (C^{\beta}_{11}-C^{\alpha}_{11})(\epsilon_{xx}+\epsilon_{yy})\epsilon^* 
      +(C^{\beta}_{12}-C^{\alpha}_{12})(\epsilon_{xx}+\epsilon_{yy})\epsilon^* \\
      &+(C^{\beta}_{11}+C^{\beta}_{12})\epsilon^{*2},			     \\
 Z_1 &= \dfrac{1}{2}(C^{\alpha}_{11}-C^{\beta}_{11})(\epsilon_{xx}+\epsilon_{yy})\epsilon^* 
      - C^{\beta}_{11}(\epsilon_{xx}+\epsilon_{yy})\epsilon^*		\\
      &+ (C^{\alpha}_{12}-C^{\beta}_{12})\epsilon_{xx}\epsilon_{yy}
      - \dfrac{1}{2}C^{\beta}_{12}(\epsilon_{xx}+\epsilon_{yy})\epsilon_{yy}\\
      &+ 2(C^{\alpha}_{44}-C^{\beta}_{44})\epsilon^2_{xy},\\
 Z_0 &= \dfrac{1}{2}(C^{\beta}_{11}(\epsilon^2_{xx}+\epsilon^2_{yy})
      + C^{\beta}_{44}\epsilon_{xy} + C^{\beta}_{12}\epsilon_{xx}\epsilon_{yy}).\\
\end{align*}

\subsection{1D stress-strain profiles}
\label{1d_stress-strain}
\begin{figure}[htbp!]
 \centering
 \includegraphics [width=0.8\linewidth]{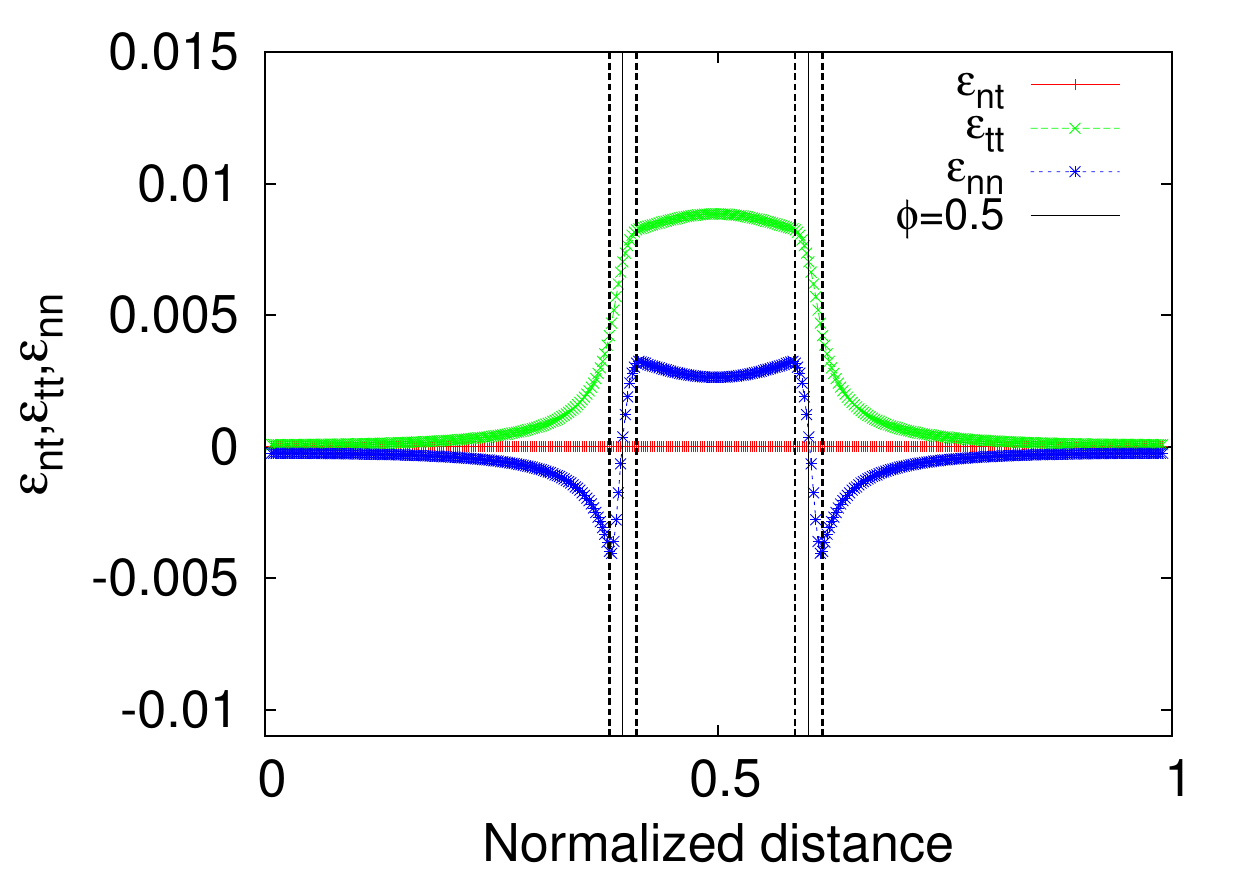} \\
 \vspace{0in}
 \caption{one dimensional strain profile across the center of precipitate where region occupied between the dotted lines 
	  represents the diffuse interface, at $\delta=0.5$, $\mu_{mat}=125$.}
 \label{1d_strain}
\end{figure}

\begin{figure}[!htbp]
 \centering
 \includegraphics [width=0.7\linewidth]{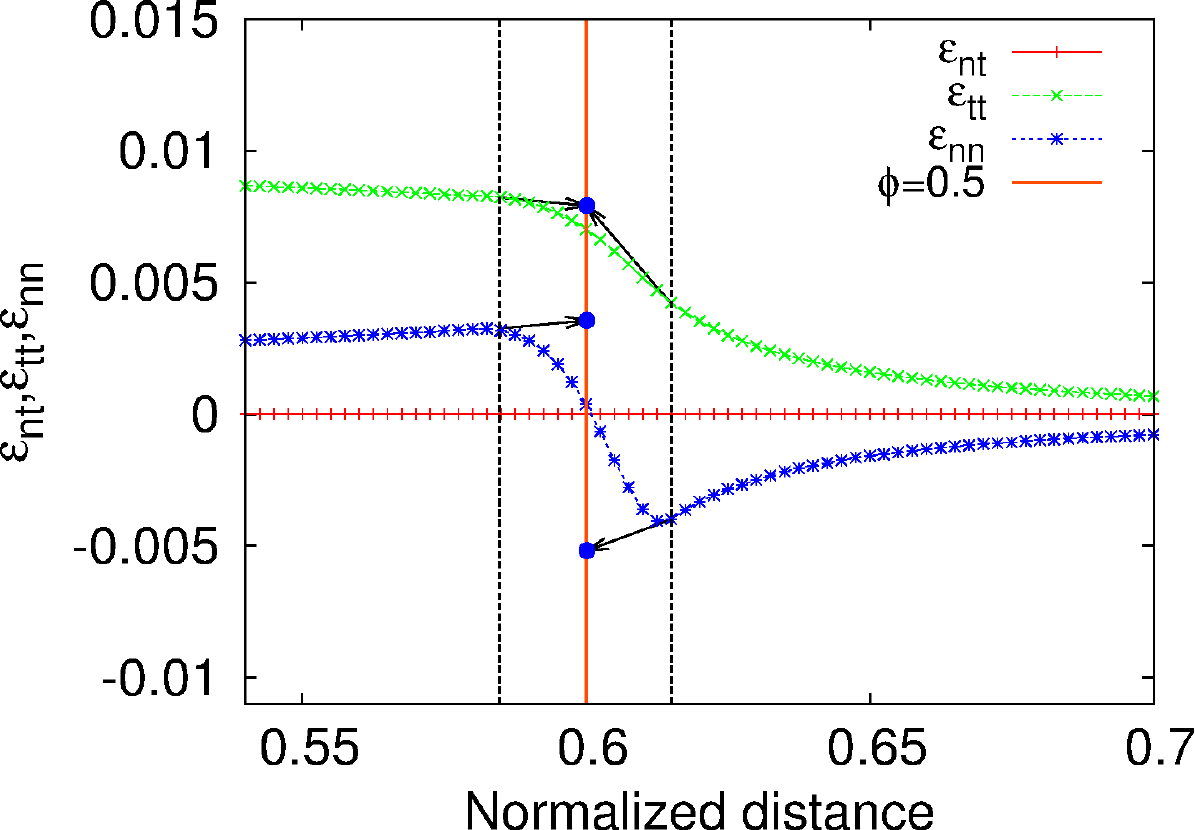} \\
 \vspace{0in}
 \caption{The zoomed section of interface showing 1D strain profile across the center of precipitate, 
	  thick orange line passes through the interface at $\phi=0.5$.}
 \label{1d_strain_inset}
\end{figure}

\begin{figure}[!htbp]
 \centering
 \includegraphics [width=0.8\linewidth]{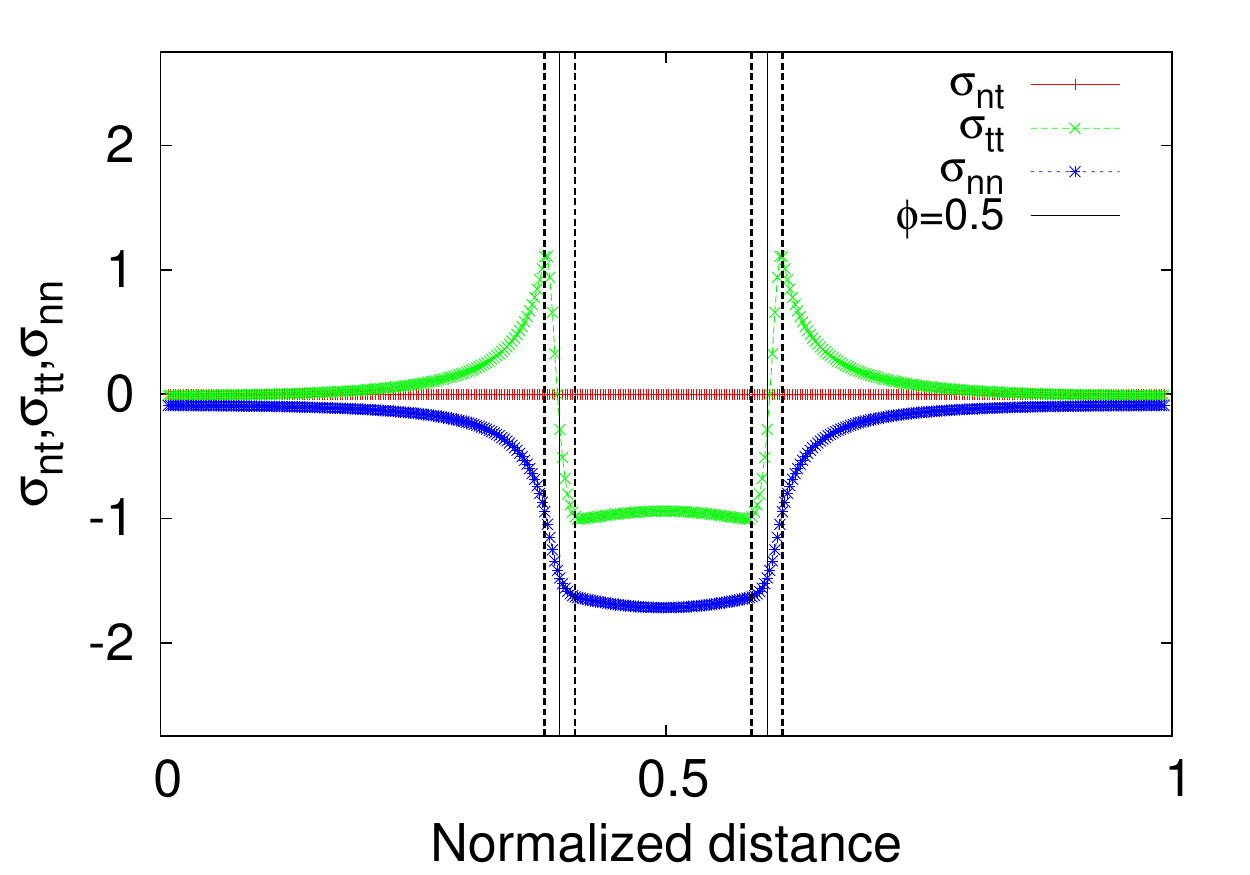} \\
 \vspace{0in}
 \caption{one dimensional stress profile across the center of precipitate where region occupied between the dotted lines 
	  represents the diffuse interface, at $\delta=0.5$, $\mu_{matrix}=125$.}
 \label{1d_stress}
\end{figure}

\begin{figure}[!htbp]
 \centering
 \includegraphics [width=0.7\linewidth]{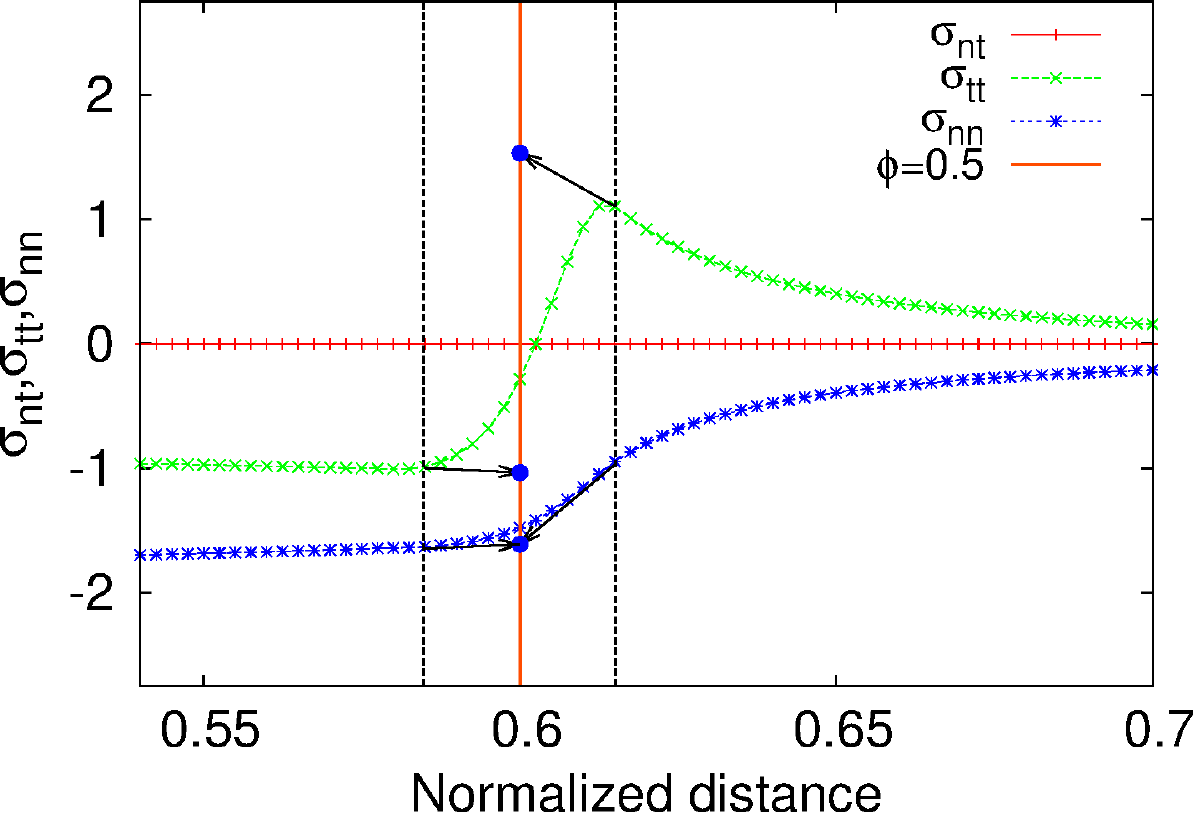} \\
 \vspace{0in}
 \caption{The zoomed section of interface showing 1D stress profile across the center of precipitate, 
	  thick orange line passes through the interface at $\phi=0.5$.}
 \label{1d_stress_inset}
\end{figure}

\section{Acknowledgements}

We thank Prof. T.A. Abinandanan for many valuable inputs and insightful, stimulating discussions on the subject.
We express our appreciation to Boeing-India for the support under the project PC36032.

\bibliography{paper_ref.bib}

\listoffigures

\end{document}